%% file: main.tex
\documentclass[sigconf]{acmart}

\AtBeginDocument{%
  }

\copyrightyear{2026}
\acmYear{2026}
\setcopyright{cc}
\setcctype{by}
\acmConference[CHI '26]{Proceedings of the 2026 CHI Conference on Human Factors in Computing Systems}{April 13--17, 2026}{Barcelona, Spain}
\acmBooktitle{Proceedings of the 2026 CHI Conference on Human Factors in Computing Systems (CHI '26), April 13--17, 2026, Barcelona, Spain}
\acmPrice{}
\acmDOI{10.1145/3772318.3791140}
\acmISBN{979-8-4007-2278-3/2026/04}

\begin{document}
\newcommand{\new}[1]{\textcolor{black}{#1}}
\newcommand{\camera}[1]{\textcolor{black}{#1}}

\author{Jian Zhang}
\affiliation{%
  \institution{University of Melbourne}
  \city{Melbourne}
  \state{VIC}
  \country{Australia}}
  \orcid{0009-0009-1619-8297}
  \email{jianzhang10@student.unimelb.edu.au}

\author{Wafa Johal}
\affiliation{%
  \institution{University of Melbourne}
  \city{Melbourne}
  \state{VIC}
  \country{Australia}
}
\orcid{0000-0001-9118-0454}
\email{wafa.johal@unimelb.edu.au}

\author{Jarrod Knibbe}
\affiliation{%
  \institution{The University of Queensland}
  \city{St Lucia}
  \state{QLD}
  \country{Australia}
 }
 \email{j.knibbe@uq.edu.au}
 \orcid{0000-0002-8844-8576}
\title{Modelling Visuo-Haptic Perception Change in Size Estimation Tasks}



\begin{abstract}

\input{section/0_Abstract}
\end{abstract}





\maketitle

\input{section/1_Introduction}
\input{section/2_RelatedWork}
\input{section/3_Methods}
\input{section/4_Results}
\input{section/5_Discussion}
\input{section/6_Conclusion}
\bibliographystyle{ACM-Reference-Format}
\bibliography{base}

\end{document}

%% file: section/0_Abstract.tex
Tangible interactions \new{involve} multiple sensory cues, enabling the accurate perception of object properties, such as size. Research has shown, however, that if we decouple these cues (for example, by altering the visual cue), then the resulting discrepancies present new opportunities for interactions. Perception over time though, not only relies on momentary sensory cues, but also on a priori beliefs about the object, implying a continuing update cycle. This cycle is poorly understood and its impact on interaction remains unknown. We study (N=80) \new{visuo-haptic perception} of size over time and (a) reveal how perception drifts, (b) examine the effects of visual priming and dead-reckoning, and (c) present a model of \new{visuo-haptic} perception as a cyclical, self-adjusting system. Our work has a direct impact on illusory perception in VR, but also sheds light on how our visual and haptic systems cooperate and diverge.

%% file: section/1_Introduction.tex
\section{Introduction}

Humans integrate \new{cues} across multiple sensory modalities in order to understand the world around them~\cite{hornbaek2025introduction}.
When you walk into a new room, visual cues give you the rough spatial layout, thermal cues let you know the heating is on, and audio cues might inform you that there is machinery around the corner. Together, this information combines into a more detailed, perceptual model of the space. 

Not only is integration important for our understanding of the world, 
but also for our ability to act within it. Visual cues can provide information about an object's location and size, for example, 
but accurate and controlled interaction also requires detailed information about weight and mass distribution (among many other things), requiring, at least, the addition of tactition and kinaesthesia ~\cite{liu2021quantitative,napier1956prehensile}. 

Research has shown that there are exciting opportunities for new interactions and experiences when designers actively \new{introduce conflicting} 
sensory cues. For example, by decoupling visual and haptic feedback, the rubber hand illusion can make participants believe that a fake rubber hand is their actual hand~\cite{botvinick1998rubber}. This works by simultaneously stroking the participant's real hand and a dislocated rubber hand, thereby matching the tactile stimuli while introducing an offset error through both the visual and proprioceptive stimuli. 
For interaction, these visuo-haptic illusions have proven popular in virtual reality, for example, to alter users' reaches through haptic retargeting~\cite{azmandian2016haptic}, or to alter their grasps through resized grasping~\cite{bergstrom2019resized}.

These perceptual illusions, however, only work within a fairly limited scope. As a result, much research effort is given to investigating and understanding perceptual limits for these interactions (e.g., \cite{azmandian2016haptic, bergstrom2019resized, jian2025}). For example, haptic retargeting illusions have been shown to work up to approx. 15\textdegree{} around an arc to the left/right of a virtual object \cite{clarence2022investigating}, and grasping illusions work up to approx. 17\% just noticeable difference for 6cm objects ~\cite{jian2025}. 

Work in sensorimotor neuroscience and cognitive psychology tells us about multiple important mechanisms of perception that come to inform these limits. For example, we know that the perceived reliability of sensory information changes \cite{ERNST2004162}, and that multi-sensory cues are integrated according to functions similar to maximum-likelihood estimators \cite{ernst2002humans}. We also know that sensory memory impacts our estimations about objects (i.e., the results of our integrated perception) through processes such as cue integration theory \cite{fetsch2013bridging}. And research has also found closed-loop processes between perception and interaction in motor adaptation ~\cite{ernst2002humans,yu2024metrics}, indicating a changing process of cognition and behaviour. 

Existing work in HCI has begun to discuss how different patterns of movement and interaction may impact the reliability of different perceptual channels (e.g., ~\cite{clarence2022investigating}), but little attention has been given to sensory integration and memorisation over time. 
Previous explorations of illusory limits were estimated either within a short time duration of use (e.g. approximately 10 minutes for each physical object in the resized grasping estimation~\cite{bergstrom2019resized}) or in an experimental context where the participants have almost no knowledge about the appearance and mechanism of the haptic proxies~\cite{bergstrom2019resized, 10.1145/3472749.3474782}, drastically impacting the available prior sensory information.
We do not know, however, how continued illusory perception (i.e., perception of mismatched multi-sensory cues) compounds -- \textit{do visuo-haptic illusions continue to work over time?} We also do not know how momentary sensory realignment impacts object perception -- \textit{what happens to an illusion when a user sees the actual proxy device?} The answers are vitally important in prolonged haptic interactions and meaningful for understanding human cognition.

\new{We study visuo-haptic perceptual conflict resolution, specifically considering the effects of \textit{time} (as acclimation, through prolonged use) and \textit{knowledge of the haptic device} (through select visual priming). We conduct one-hour user studies in VR, during which we vary either the visual cue (with a passive haptic device) or both the visual and haptic cues (with an active, shape-changing device). }
We designed a device that dynamically alters its size between 6cm and 8cm. We ran multiple
forced choice tasks to estimate the perception, alongside acclimation games to simulate continued exposure and practice with the device.
We controlled the participant's view of the device, either never revealing it (non-priming), or revealing the true device (true visual priming) or a false device of a different size (misleading priming) at certain times. 
We recruited 80 participants and divided them into 4 groups with conditions of (1) non-priming fixed-size device, (2) non-priming size-changing device, (3) fixed-size device with correct visual priming, and (4) fixed-size device with a misleading visual priming.


Based on our empirical results, we make the following claims:
\begin{itemize}
    \item Perception drifts over time \camera{and the size is increasingly overestimated for passive objects,} following a curve to a local asymptote
    \item Active, shape-changing devices skew subsequent size perception differently to passive objects. \camera{The perceived size drifts to be underestimated when the physical size actively shrinks.}
    \item Visual priming \camera{sets a more accurate initial perception and} reduces perceptual drift across a whole experience
    \item Intermittent visual priming serves as a form of dead-reckoning, resetting perception towards that prime
    \item Sensory integration can be modelled as a first-order control system, wherein the perceptive discrimination works as an amplifier and the feedback loop is influenced by confidence in the perception
\end{itemize}

These claims (as contributions) have important implications for tangible interactions. First, they begin to describe what we should expect if we deploy visuo-haptic illusions, such as resized grasping~\cite{bergstrom2019resized}, over time. Second, they reveal how selective priming and dead-reckoning may enable designers to skew or reset perception at key moments. And third, they provide early insights into what users come to believe about tangible interactive objects when their vision is directed away from it.

%% file: section/2_RelatedWork.tex
\section{Related Work}
\camera{Mismatched visual and haptic cues can easily occur in scenarios with screens and head-mounted displays. 
This can happen accidentally, such as a result of tracking errors and latency causing spatial misalignments between visual cues and their physical, haptic counterparts. But it can also be leveraged purposefully, through the design of specific haptic devices and visuals, that induce mismatches for novel experiences and opportunities. Integral to this opportunity, however, is an understanding of how visuo-haptic cues are perceived and integrated by users.  }

We review recent literature on haptic devices and highlight how visuo-haptic illusions are being leveraged and understood.

 

\subsection{\camera{Haptic Devices}}
Efforts to deliver high-resolution haptic feedback have led to the development of a variety of tangible devices targeting immersive interaction, especially in VR. Recent advances have largely concentrated on wearable and handheld formats. Representative examples include wearable systems for grasping rigid forms~\cite{choi2016wolverine, choi2018claw}, objects changing geometry features~\cite{10.1145/3472749.3474782}, and complex shapes via multi degree-of-freedom mechanisms~\cite{ulan2024}. However, these devices often face challenges of being complex, cumbersome, and costly.

By contrast, grounded haptic systems (particularly those of encounter-type interfaces) can offer feedback only when necessary, leaving users hands free. Commercial products such as the Touch\footnote{3D Systems, USA – originally developed by SensAble Technologies} and Omega\footnote{Force Dimension, Switzerland}, as well as research prototypes like inFORM~\cite{follmer2013inform}, ShapeShift~\cite{siu2018shapeshift}, and REACH+\cite{gonzalez2020reach+}, are examples of such devices enabling different interactions. These platforms are capable of delivering highly detailed feedback but similar to the wearable devices, their complexity, cost, and infrastructural requirements limit their accessibility for everyday interaction.

As a result, research has also been exploring passive haptic techniques with simple designs or daily accessible objects. 
For example, by leveraging only simple geometric building blocks, HapTwist demonstrated a range of complex haptic objects 
~\cite{10.1145/3290605.3300923}). 
Another popular avenue has been to use everyday objects, in concert with visuo-haptic illusions to, e.g., haptically render objects of different sizes~\cite{bergstrom2019resized} and in different locations~\cite{azmandian2016haptic}. 


\subsection{Visuo-Haptic Illusions}

Usually, visual and haptic stimuli match and integrate with each other to form a perception of reality in interactions such as touching and grasping. In scenarios such as extended reality, however, tangible interaction often features a decouple and recouple process of the visual and haptic cues by blocking the real vision and rendering a virtual one. These visuo-haptic discrepancies, if tightly controlled, can create new experiential and interactive opportunities. The rubber hand illusion~\cite{botvinick1998rubber}, for example, introduces a spatial discrepancy between the user's real hand and a fake rubber hand that, combined with synchronous visual and tactile cues, allows the user to perceive body-ownership of the fake hand.

Research in this space continues to develop and expand the boundaries of these illusions. Much of the work on visuo-haptic illusions has focused on two primary purposes: illusions of 
\textit{where} the object is, or \textit{what} the object is. 

Illusions of \textit{where} have been heavily influenced by redirected touching~\cite{kohli2010redirected} and haptic retargeting~\cite{azmandian2016haptic}. These illusions seek to guide the user's physical hand towards a proxy object that is spatially decoupled from its virtual counterpart. Examples have demonstrated redirect controller buttons~\cite{zenner2019estimating,gonzalez2019investigating}, enabled users to grab objects placed around them~\cite{clarence2022investigating}, and attempted to retarget random, unscripted reaches~\cite{clarence2021unscripted}. From work on these illusions, we have come to understand the spatial \textit{haptic coverage} of a physical object -- the area within which it can provide haptic feedback for virtual objects~\cite{clarence2024stacked}.


Illusions of \textit{what} the object is largely explore the extent to which one physical object can feel like another. In addition to the devices that can directly alter the physical size stimuli (e.g. CLAW~\cite{choi2018claw}, X-Rings~\cite{10.1145/3472749.3474782}, etc.), applications of illusions aim to convince users they are interacting with an object with one property (for example, a heavy hammer), while they in fact interact with an object with a different property or, at least, with an object with a different magnitude of that property (e.g., a lightweight bottle). 

In a visuo-haptic system, simple physical devices can express various shapes with edges, curves, and surfaces~\cite{ban2012modifying, ban2014displaying}. The method was further developed by Zhao et al.~\cite{zhao2018functional} to extend haptic retargeting to complex, arbitrary shapes. Bickmann et al.~\cite{bickmann2019haptic} showed that haptic feedback can be created by illusion without grasping any physical prop and developed a haptic illusion glove that can provide haptic feedback for various virtual objects such as a cup, a hammer, and a water can. Yang et al.~\cite{yang2018vr} demonstrated that the illusion of size change could be induced not only through direct hand interaction, but also when manipulating virtual tools such as chopsticks. In many of these applications, the illusion-based tangible interaction often relies on visual distortion to alter perception. Kim et al.~\cite{kim2024big} introduced a fixed-size haptic controller that uses finger repositioning to create the illusion of dynamic size change. This allows users to perceive changes in object size with proper visual feedback. Furthermore, illusions have been used to simulate factors in interaction such as geometry ~\cite{turchet2010influence}, weight~\cite{Zhang2025ThumbShift}, force feedback~\cite{lecuyer2000pseudo}, stiffness ~\cite{sanz2013elastic} and texture ~\cite{brahimaj2023cross}. The research community has developed the approaches described above to either induce or exploit visuo-haptic illusions. Consequently, the underlying mechanisms and characteristics of these illusions (e.g., their limits) have also attracted much attention.


\subsection{\camera{The Mechanism of Visuo-haptic Illusions and Adaptation}}

Modern theories describe our cognition and understanding of the world around us as a predictive process. That is, our brains make predictions about the world around us, based on expectations, and our previous experience and knowledge. These predictions are then compared to the information we receive through sensory cues, and subsequently updated and corrected in a closed-loop adaptation process~\cite{clark2024experience, hornbaek2025introduction}. 

\camera{In cognitive science, research explains this adaptation as an active process where humans adjust their perception and behaviour to improve each other. Sensory inputs, while being noisy or unreliable sometimes, are processed by the brain in a fashion similar to a maximum-likelihood integrator, which can variously weight and attenuate different cues to update our perception~\cite{ernst2002humans,kording2004bayesian, berniker2011bayesian}.} 

\new{Based on this adaptation and integration of multi-sensory cues, several studies have attempted to model the patterns of the visuo-haptic illusions and accordingly proposed hypothetical theories. For the size perception, \citet{9580906} has described a simple linear pattern of perceived size increasing with the virtual size in augmented reality. Similarly, \citet{bergstrom2019resized} built a model indicating that illusions are more easily detected when objects are larger.
\citet{jian2025} further built a multi-dimensional model of the interaction of size and angle in visuo-haptic perception. These models focused on the impacts of the physical properties and human biological features (e.g. hand sizes) on the visuo-haptic illusions.} 

\camera{Furthermore, more complex models have been proposed in related tasks such as haptic retargeting. \citet{10.1145/3491102.3501907} recently proposed a control process with an internal motion controller updating the hand state in a closed-loop system in haptic retargeting. Based on this model, \citet{lebrun:tel-04088544} further adapted a stochastic optimal feedback control model to describe the relationship between natural tendency with illusions when performing haptic retargeting. These models apply control theories in human perception and motor control, attempting to interpret the visuo-haptic perception from the perspective of cognitive science and neurobiology.}

Many of these models consider visuo-haptic illusions and integration as static. As these cues are prone to error and noise, however, this updating process is continuous ~\cite{jeschke2024humans}, and
human activities, in general, involve a large amount of learning and adaptation over time, such as to improve motor control~\cite{krakauer2019motor}. Similar to a control system with feedback loops, errors are corrected and the larger the errors are, the larger the corrections are (e.g., in motor learning). As part of motor control, perception based on multisensory inputs plays an important role in the learning process. Specifically in tangible interactions, cases such as rubber hand illusion~\cite{botvinick1998rubber} have shown the effect of an acclimation process in coupling sensory stimulus and creating the illusions.

Previous studies indicate the existence of an adaptation process in motor control and perception~\cite{johansson1992sensory, jeschke2024humans}. While the process of changing perception has been influencing experiences and immersion in tangible interactions, this cognitive process of fundamental properties such as size perception, however, has not been estimated or explored.

%% file: section/3_Methods.tex
\section{Methods}

We ran a study to examine the impact of time and visual priming on visuo-haptic illusions. Specifically, through four conditions, we estimated
(1) the impact of prolonged interaction on perception of fixed-size objects, (2) the impact of prolonged interaction on perception of size-changing devices, (3) the impact of correct visual priming and (4) the impact of incorrect visual priming on size perception. We designed a three-phase study, including (a) estimation tasks, (b) breaks, and (c) an acclimation game. 
The estimation task captured participants' size perception through a \new{forced choice task, where participants were presented with haptic and/or visual stimuli and asked to choose between two alternative size options} in VR. The break phase included filling out a NASA Task Load Index questionnaire on paper. The acclimation game, in VR, was designed to ensure prolonged haptic exposure to the device, making participants grasp, hold, and release the device repeatedly.

\subsection{Apparatus}
\subsubsection{Haptic Proxy Design}
\begin{figure*}
  \includegraphics[width=0.95\textwidth]{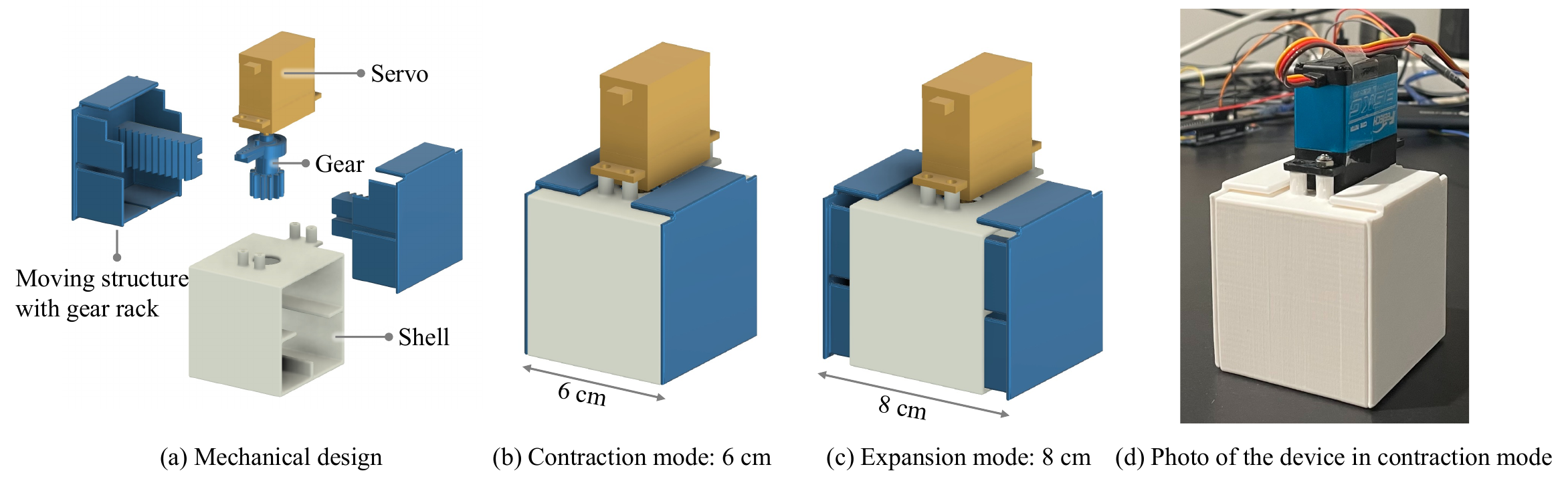}
  \caption{A device changes its size (width of grasp) between 6 cm and 8 cm. (a) Breakdown mechanical design of the device. The motor and gear in the centre drives the sides of the cube via the gear racks to expand/contract. (b) The contraction mode of the device with the appearance of a cube with the width of 6 cm. (c) The expansion mode of the device with the width of 8 cm. (d) The photo of the prototype (in contraction mode).}
  \Description{The figure presents a device capable of changing its grasp width between 6 cm and 8 cm using a servo and gear-driven mechanism. The mechanical design (a) shows the servo mounted on top and connected to a blue circular gear beneath it. This gear interfaces with moving structures containing gear racks that slide to expand or contract the cube’s sides. The entire assembly is supported by a shell, which provides the structural housing for the mechanism. In contraction mode (b), the cube takes on a compact form with a width of 6 cm. The sliding panels are drawn inward, giving the appearance of a neat cube. In expansion mode (c), the sliding panels are driven outward, increasing the device’s width to 8 cm while retaining its cubic form. The prototype photo shows the real-world version of the device in contraction mode at 6 cm.}
  \label{fig:devicedesign}
\end{figure*}
We prototyped a device that can change its size (specifically, the width for grasping) between 6 cm and 8 cm, as shown in Figure \ref{fig:devicedesign}. An Arduino UNO board controls a FT5330M aluminium servo to rotate the gear in the centre of the device, and drives the gear racks to expand/contract the sides of the cube. This servo offers up to 35$kg/cm$ of torque, ensuring that the cube will achieve the specified size (i.e., it is not easily overpowered by the participant). It takes 0.9$\pm$0.1 seconds to change size between 6 cm and 8 cm with the error less than 0.05 cm, and the speed change when applying a grasp force on the device is negligible according to technique estimation.

The device weighs 162$\pm$1 grams (including Optitrack markers). The device was fully 3D printed with polyethylene terephthalate glycol
(PETG), except for the servo, electronic components, and fastening hardware. The device was connected with long jumper wires
(approximately 40 cm), to ensure a free, unrestricted movement. When contracted, the device is a 6.0$\pm$0.05 cm cube.

\subsubsection{Virtual Scene and Tracking System}
\begin{figure*}
  \includegraphics[width=0.95\textwidth]{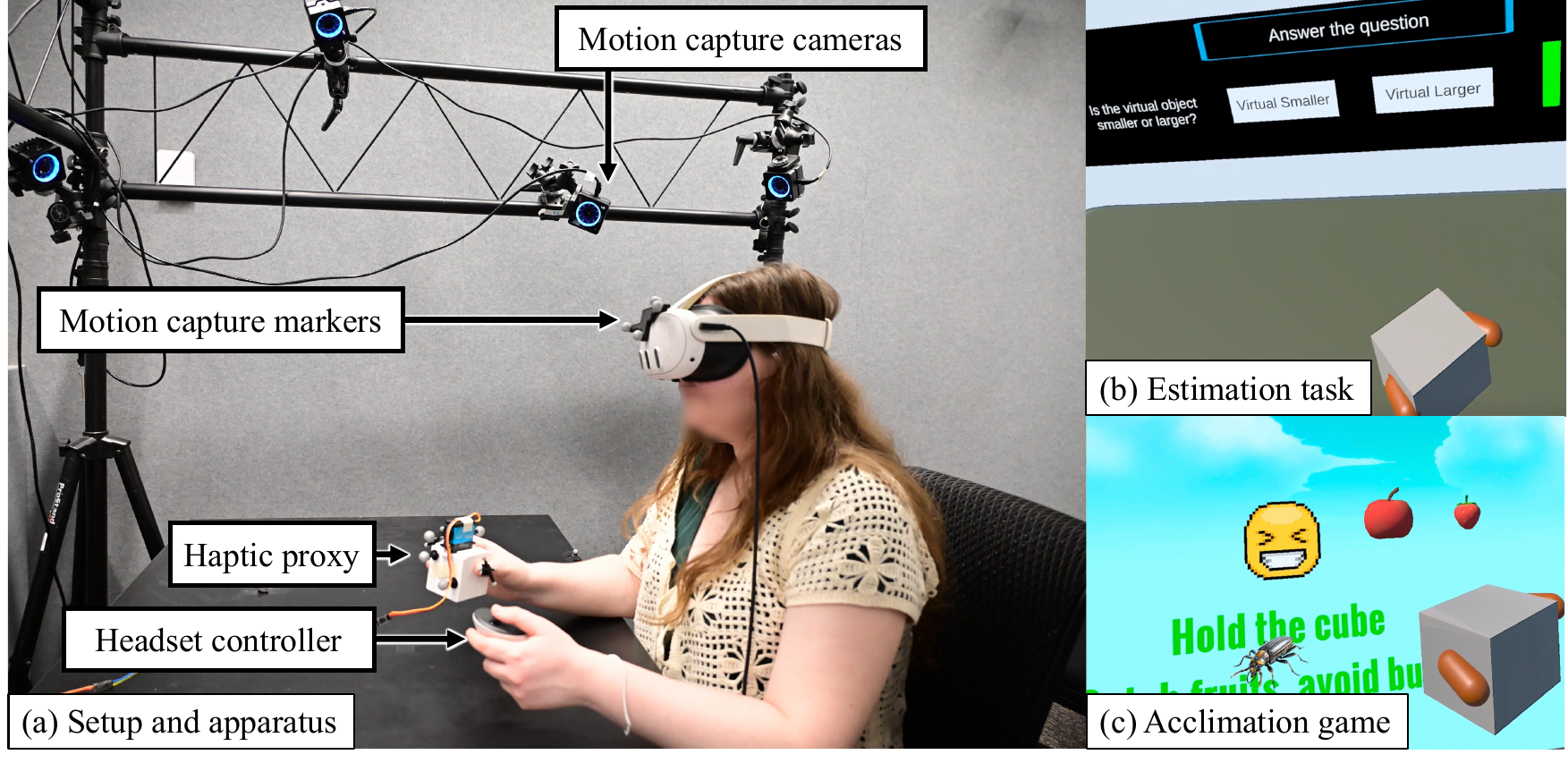}
  \caption{Experimental setup. (a) shows the apparatus and overall setup of the physical environment. (b) and (c) are accordingly the virtual scenes of the estimation task and acclimation game.}
  \Description{The figure presents both the physical apparatus and the virtual environments. In (a), the setup and apparatus are displayed with a participant seated at a table wearing a VR headset. Motion capture cameras are arranged around the participant to track movement, and reflective motion capture markers are attached to the environment and equipment for precise tracking. The participant holds a haptic proxy designed to simulate interaction with virtual items, while also using a headset controller. This arrangement allows integration of physical feedback with virtual tasks. Figure (b) shows the virtual scene for the estimation task. On the VR display, participants are asked whether the virtual object appears smaller or larger compared to the physical one. Figure (c) shows the virtual acclimation game. The screen instructs the participant to hold the cube and avoid some objects and catch the others.}
  \label{fig:settings}
\end{figure*}

The VR scene was built in Unity 2022 on a laptop PC (13th Gen Intel i9-13900HK 2.60GHz, 32.0GB RAM, NVIDIA GeForce RTX 4090 GPU, Windows 11 Pro). The scene was presented to the participants via a Meta Quest 3 head-mounted device (HMD). We used an Optitrack motion capture system for tracking the device, the hand movement, and the headset. There were 7 Prime 13W cameras, running at 240Hz, placed around the table where the participants performed the tasks (see Figure~\ref{fig:settings}). The markers of the motion capture system were mounted on the participant's thumb and index finger, the device for grasping, the HMD, and the table. Markers mounted on the tracked objects did not interfere with the participants' range of motion or object interactions. 

\subsubsection{Finger Tips Rendering}
Although only the thumb and index finger of the participant's right hand were tracked and rendered with capsule-shaped objects in the virtual scene, they were asked to hold the device with all the fingers of their right hand, also known as Large Diameter grasp in the GRASP taxonomy~\cite{feix2015grasp}. \new{As our participants were grasping parallel faces of a cube, we assume their fingers are in a line opposing their thumb. To prevent tracking marker occlusion and to reduce confounding effects of hand rendering (e.g., ~\cite{jorg2020virtual,10049656}), we use only a simplified and reliable rendering scheme to assist with targeting.}

Similar to previous studies \cite{jian2025,bergstrom2019resized}, we applied a resized grasping algorithm between the fingers in real-time. The distance of the index finger and thumb to be rendered in the virtual scene $D_{v}$ was calculated based on the scaling ratio between the widths of the virtual object and the physical device $S_{v}/S_{p}$, where $S_{v}$ is the size of the virtual object and $S_{p}$ is the size of the physical object  (note that the virtual object size does not always equal the physical object size, as mismatching sizes were presented for the estimation tasks and for a random visual stimuli in the acclimation game). The distance $D_{v}$ between the virtual index finger and thumb was resized by the method:
\small
\begin{equation}
    D_{v} = D_{p}\times(S_{v}/S_{p})
\end{equation}
\normalsize

where $D_{p}$ is the distance between the index finger and the thumb in the real world. Through resizing, the participant's fingers were always aligned to the virtual cube when they grasped the physical device. 

\subsection{Experimental Procedure}

\begin{figure*}
  \includegraphics[width=0.95\textwidth]{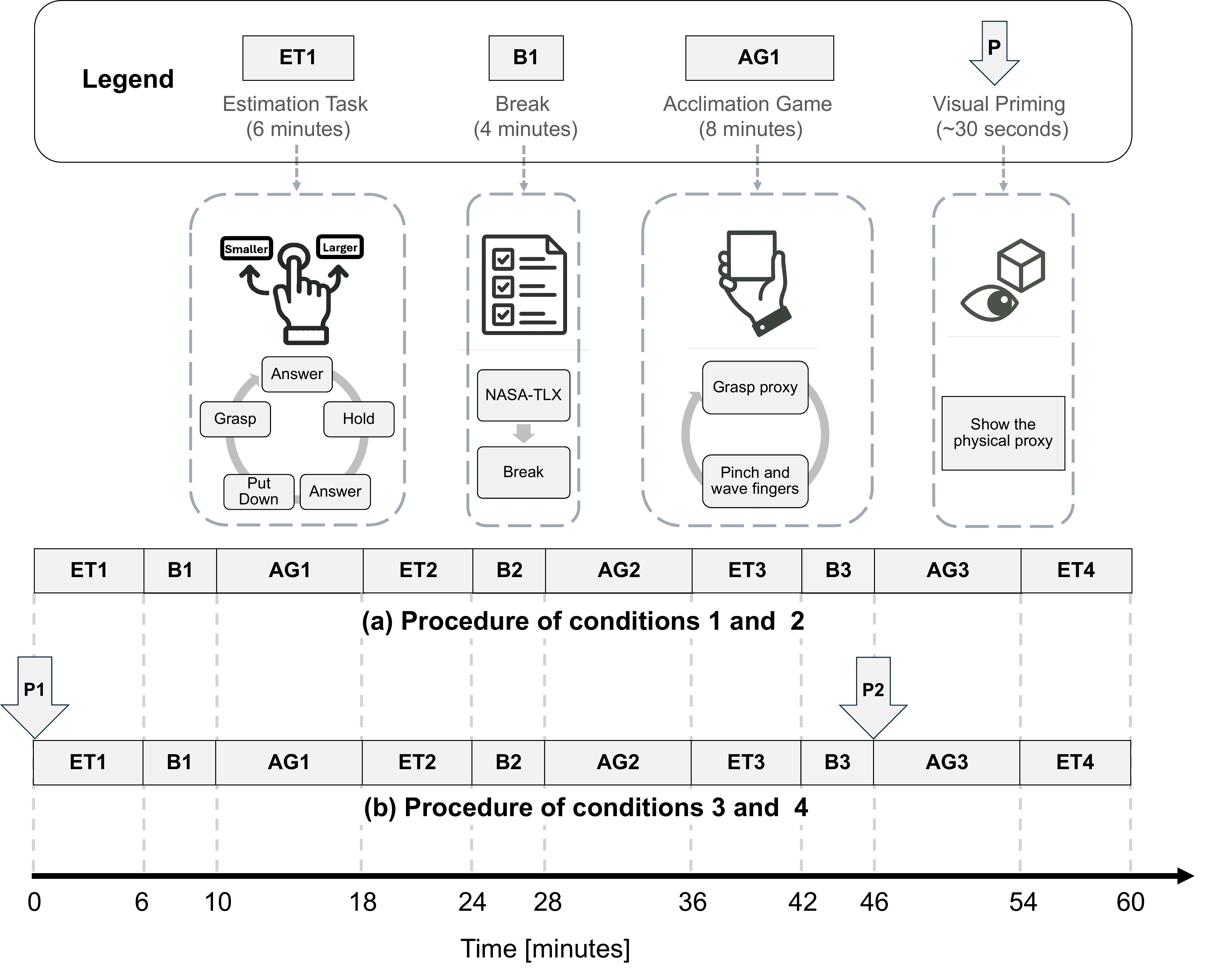}
  \caption{The procedure includes four 6-minute estimation task blocks (an initial one and three after each acclimation game block), three 4-minute break blocks (before each acclimation game block) and three 8-minute acclimation game blocks. The overall designed study time is 60 minutes. The figure shows (a) procedure of conditions 1 and 2 without any priming and (b) procedure of condition 3 and 4 with visual priming at the beginning of the study and the beginning of the last (third) acclimation game block.}
  \Description{The figure illustrates the experimental procedure for a 60-minute study consisting of estimation tasks, breaks, acclimation games, and visual priming. At the top, diagrams provide a summary of each activity. The estimation task involves participants making size forced choice answers (smaller or larger) after grasping and when holding a device object, and then put it down for later grasping again. The break block contains completing the NASA-TLX workload questionnaire and resting. The acclimation game block requires participants to interact with a grasp proxy, also including pinching and waving their fingers, training them to get used to the device. The visual priming block shows participants the physical device object. Timeline (a) shows the procedure for conditions 1 and 2, where no priming is used. The schedule begins with ET1, followed by B1, AG1, ET2, B2, AG2, ET3, B3, AG3, and finally ET4, covering the 60-minute period. Timeline (b) shows the procedure for conditions 3 and 4, which include visual priming. Here, priming (P1) occurs before the first estimation task and priming (P2) occurs again before the final acclimation game block.}
  \label{fig:procedure}
\end{figure*}

The procedure is shown in \autoref{fig:procedure}. To evaluate the influence of time and visual priming on both fixed-size  (\textit{passive}) devices and size-changing (\textit{active}) devices, the study was designed into 4 conditions. 

\subsubsection{Estimation Task (6 minutes)}
The Estimation Task (ET) was carried out by presenting forced choice questions to the participants and collecting their answers, similar to the task settings in previous studies~\cite{bergstrom2019resized, de2019different, jian2025}. The participants were asked to grasp the device with all the fingers of their right hands, compare the physical size in their right hands and the virtual size they saw in the VR scene, and choose between ``virtual smaller'' (the virtual size is smaller than the physical size they perceive in the hands) and ``virtual larger'' (the virtual size is larger than the physical size they perceive in the hands) within 6 seconds (visually indicated by a time bar to the participants). They input their answer with the Meta Quest 3 controller in their left hand. After answering the question, the participants were asked to hold the device in the air and wait for 2 seconds. During this wait, the virtual fingers and objects were all invisible, and, in the size-changing condition, the device would change size. At this point, the virtual object and the fingers would become visible again.
Subsequently, while still holding the device, the participants then compared a new virtual object in their hands with the physical one. After answering this second question, the participants were asked to put the device down on the table, rest their hand in front of their right side, 
and wait for the next group of questions. During this period, the virtual object was not visible. 
Throughout the whole study, participants would lift - answer - wait - answer - put down. This allowed us to examine the immediate impact of size-change (during the wait period) on perception. We use the same procedure when the physical size does not change, to avoid introducing time confounds between conditions. 

Throughout the estimation block, participants compared physical objects against virtual objects between $\pm{}$ 2.0 cm in 0.5 cm steps (e.g., the 6 cm physical object was compared against nine virtual objects between 4.0 cm and 8.0cm, and the 8 cm physical object was compared against nine virtual objects between 6 cm and 10 cm). 
In the size-changing condition, the physical object would change from 6 cm to 8 cm (expand) and from 8 cm to 6 cm (contract). Each physical size was compared against the nine virtual sizes, for 18 total questions. In the fixed-size condition, always 6.0 cm, to keep the duration the same, each virtual size was presented twice (1 size $\times$ 9 virtual sizes $\times$ 2 times = 18 questions). 

\new{In the fixed-size condition, we only analyse the results of the first question in each lift, to understand the impact of grasp on size perception. We believe the second response is likely biased by the first answer. Additionally, as the device is not released and grasped again, the participants receive less haptic information for consideration in their second response (in contrast to receiving new, conflicting visual information). As such, we chose not to analyse the data of the fixed-size second response. We keep the lift-respond-wait-respond format, however, to standardise task duration for considering acclimation. For transparency, we include the omitted results in the supplementary material.}

\new{As for the size-changing cube, the second answer was analysed to simulate a size-changing device performing changes in the users’ hands. Although there was no “putting down” action, the haptic stimuli and visual stimuli both changed, which could serve as a reset and minimise the impact of the former question.}

The order of presentation of the virtual sizes was randomised.
The virtual sizes were derived from the results of prior work~\cite{bergstrom2019resized}.


In the \textbf{size-changing} condition, 
the order of the size changes(6 cm-8 cm and 8 cm-6 cm) was also randomised.

For each question, the participants had 6 seconds to answer the question while lifting the device. The data would be invalid if they failed to answer within the time limit.

\subsubsection{Break (4 minutes)}
During every break a NASA Task Load Index questionnaire~\cite{hart1988development} was filled out by the participants.
There were 21 gradations on the scales with 0 - ``Very Low'' and 20 - ``Very High'', except the fourth question where 0 means ``Perfect'' and 20 means ``Failure''.

\subsubsection{Acclimation Game (8 minutes)}

We designed a short acclimation game for the participants to complete with the study device. Participants used the device to move a virtual object in the scene.
The game lasted for 8 minutes with 8 levels (1 level per minute). For the first 50 seconds of the game, the participants were instructed to ``catch the fruits dropping from the sky with the controller in their hand, and avoid catching falling bugs''. For the last 10 seconds of each level, participants were asked to ``put the device down on the table and pinch the falling bugs with their thumb and index finger''. This game was designed to both acclimate participants to the device and refresh their proprioception through different hand poses. 

During the game, the virtual object changed size randomly every 10 seconds. 
For the \textbf{fixed-size} condition, the virtual size changed randomly between 4.0 cm and 8.0cm, with a 0.5 cm step -- the same as during the estimation task. 
In the \textbf{size-changing} condition, the virtual cube changed size between 4.0 cm and 10.0cm, at 0.5 cm steps. When the virtual cube was $\leq$7.0 cm, the physical cube was 6 cm, changing to 8 cm for virtual objects larger than 7.0 cm. 


\subsection{Experimental Conditions}

We designed 4 different conditions as shown in Table \ref{tab:conditions}. \new{The participants were divided into 4 groups and the study followed a between-subject design.}

\begin{table*}
\caption{User study of four conditions}
  \label{tab:conditions}

  \begin{tabular}{c|c|c|c}
  
    \hline
    &No priming&Correct priming (6 cm)&Misleading priming (5 cm)\\
    \hline
    Fixed-size proxy&\textbf{Condition 1}&\textbf{Condition 3}&\textbf{Condition 4}\\
    \hline
    Size-changing proxy&\textbf{Condition 2}&-&-\\
    \hline
\end{tabular}
\Description{For the fixed-size proxy, three conditions were tested. With no priming, participants were in Condition 1. With correct priming at 6 cm, they were in Condition 3. With misleading priming at 5 cm, they were in Condition 4. For the size-changing device, only one condition was tested: Condition 2, which used no priming.}

\end{table*}

\subsubsection{Condition 1 - fixed-size device}
Condition 1 was the simplest condition: a \textbf{fixed-size} device of 6 cm without any visual priming. In this setup, participants never saw the physical device during the entire study, allowing us to investigate the influence of interaction time on the visuo-haptic perception in VR. As shown in Fig. \ref{fig:procedure} (a), size perception was estimated four times over the course of the 1-hour session. 

\subsubsection{Condition 2 - size-changing device}
In Condition 2, the physical device performed a \textbf{size change} in the user's hand to simulate active haptic devices in VR. The overall procedure was identical to Condition 1 (see Fig. \ref{fig:procedure} (a)). The participants never saw the physical device.

At the beginning of each estimation trial, the device was set randomly to either 6 cm or 8 cm for the participant to grasp. After the first question was answered, the device changed size while still held in the hand and a \new{forced choice} question was asked again. During the size change, and in the brief intervals between the paired questions (when the device was put down and lifted again), the virtual object was not displayed to prevent additional visual cues. Therefore, the data collected includes both perception of the object after the initial grasp and the perception after a size change.

\subsubsection{Condition 3 - correct visual prime}
In condition 1 and 2, participants were asked to finish the task without seeing the physical device. The visual priming of the physical device 
may pre-inform the perception of size, however~\cite{kim2024big}. As shown in Fig. \ref{fig:procedure} (b), right before the first estimation task and before the third (final) acclimation game, the physical device was shown in contracted mode (6 cm) to the participants as instructions were given (the instructions were repeated before the final acclimation game in condition 3 and 4). The participants didn't touch the device during the visual priming. In this condition, the physical device never changed size.

\subsubsection{Condition 4 - misleading visual prime}
To explore whether a misleading visual prime can consistently skew size perception, in condition 3 we showed participants a physical device of a smaller size (5 cm). This device was produced with a corresponding motor and cable, such as to convince the participant it was the true device. 
This device was
shown to the participants during the instructions before estimation task 1 and acclimation game 3, but was replaced by the researchers after the participants put on the headset, such that they would not notice the change of device. At no stage could they see the real physical device. 
The participants did not touch the device during the visual priming. In this condition, the device never changed size.

Across all conditions, the same device was used, for consistent weight and mass distribution.

\subsection{Participants}
We recruited a total of 80 right-handed participants for the user studies through a public university website, social media and other networks. Of these participants, 44 self-identified as female and 36 as male.

Participants ranged in age from 19 to 39, with a mean age of 25.46 (SD = 4.18). Regarding prior experience with virtual reality (VR), 31 participants reported having no experience, 41 reported some experience, and 8 reported good experience.

The study was approved by the local ethics board. Each participant received a gift card for participating in the study. 

Participants were randomly assigned across the 4 study conditions, with 20 assigned to each condition.

%% file: section/4_Results.tex
\section{Results}\label{Results}


\subsection{Condition 1 Results}
For every estimation task the proportion of answering ``virtual smaller'' among all valid answers (``virtual smaller'' and ``virtual larger'') was calculated for each virtual size. For example with the 6 cm devicen the proportion was very high for the 4 cm virtual size (because most of the participants felt the virtual one was smaller), and was very low for the 8 cm virtual size (because most of participants felt the virtual one was larger). The proportions of different virtual sizes were then fitted to a sigmoid function, following the procedure of prior work ~\cite{bergstrom2019resized, jian2025} for forced choice task diagrams:

\begin{equation}
f(x)=\frac{1}{1+e^{ax+b}}
\label{eq:sigmoid}
\end{equation}

where \textit{f(x)} is the proportion of choosing ``virtual smaller'' and \textit{x} is the virtual size. The parameters \textit{a} and \textit{b} were obtained by fitting.

\new{Thus, each data point represents the responses from 20 participants by calculating the percentage of selecting "virtual smaller" (and/or selecting "virtual larger", which won't affect the results in the study). They were analysed later in groups of 9 comparisons (9 different virtual sizes). The results estimated represent a population visuo-haptic perception instead of individual perception, following traditional HCI practices~\cite{bergstrom2019resized, jian2025}.}

There were in total 17 (among 20 participants $\times$ 9 virtual sizes $\times$ 2 times $\times$ 4 estimation tasks = 1440 questions, 1.18\%) missing answers for condition 1 because the participants did not answer the question in time.

\begin{figure*}
  \includegraphics[width=0.98\textwidth]{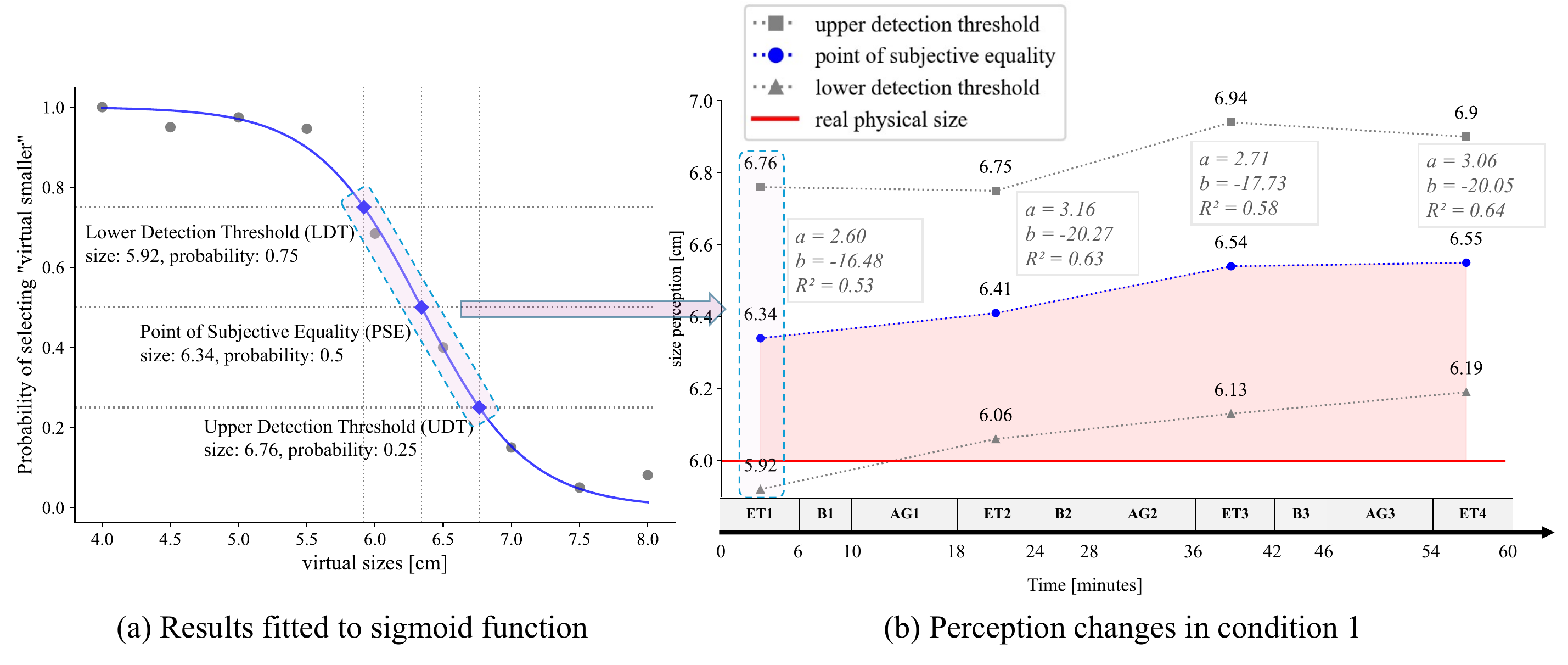}
  \caption{Data analysis of condition 1. (a) showcases an example of how the proportions of answers are fitted to sigmoid function and how \camera{upper detection thresholds}, point of subjective equality and \camera{lower detection threshold} are calculated with data of condition 1, estimation task 1. (b) plots the perception thresholds and PSE of all estimation tasks in condition 1 together with time as the horizontal axis. Fitting parameters including a, b and McFadden $R^2$ are shown alongside.}
  \Description{In figure (a), the results are fitted to a sigmoid function. The horizontal represents virtual sizes in centimetres, while the vertical represents the probability of selecting “virtual smaller.” The fitted curve shows how response probabilities vary with stimulus size. Key perceptual metrics are marked: the \camera{lower detection threshold (LDT)} at 5.92 cm with a probability of 0.75, the point of subjective equality (PSE) at 6.34 cm with a probability of 0.5, and the \camera{upper detection thresholds (UDT)} at 6.76 cm with a probability of 0.25. In Figure (b), perception changes across the duration of Condition 1 are plotted. The horizontal represents time in minutes across the sequence of blocks (ET1, B1, AG1, ET2, B2, AG2, ET3, B3, AG3, ET4), while the vertical represents size perception in centimetres. Over time, perception values fluctuate, with PSEs starting at 6.41 cm in ET1, shifting to values such as 6.54 cm in ET2 and stabilizing around 6.55 cm in ET4. The shaded region between thresholds indicates the perceptual range. Model fitting parameters for each estimation block are also shown, including slope (a), intercept (b), and McFadden’s R square, reflecting the goodness of fit.}
  \label{fig:c1}
\end{figure*}

After fitting the data to sigmoid functions of each estimation task, as shown in Fig. \ref{fig:c1} (a), we calculate the point of subjective equality (PSE, where there's a 50\% probability of selecting "virtual smaller", indicating a random choice and, thus, uncertainty about the stimuli) as well as the perception thresholds (where there's a 25\% or 75\% of selecting ``virtual smaller'', meaning the users start to be uncertain about the size, as interpreted in previous studies \cite{bergstrom2019resized, de2019different, jian2025}). The results of condition 1 are shown in Fig. \ref{fig:c1} (b), representing the PSE and thresholds as the perception in the middle of each estimation task (i.e. the first data point is considered to be at the moment of the middle of estimation task 1, 3 minutes from the beginning of the study).  \new{In this task with only 2 alternatives ("virtual smaller" and "virtual larger"), the calculated perceptual thresholds and PSEs are the same using the percentage of either "virtual smaller" or "virtual larger".}

The \camera{upper and lower detection} thresholds form an area where the users don't notice the physical and virtual size differences, while the PSE represent the absolute size perception. Both the area and PSE show a clear increasing pattern over time in condition 1, slowing down at the last estimation task (ET4). This indicates that participants perceived the physical device to be larger and larger over time, resulting in a 2.1 mm difference between the first estimation task (6.34 cm) and the last estimation task (6.55 cm). McFadden's pseudo-$R^2$ for every curve shows a very strong fit (>0.5).

\subsection{Condition 2 Results}

In condition 2, participants interacted with a size-changing device that alternated between 6 cm and 8 cm throughout the 1-hour experiment. At no point were they shown the physical device.

\begin{figure*}
  \includegraphics[width=0.95\textwidth]{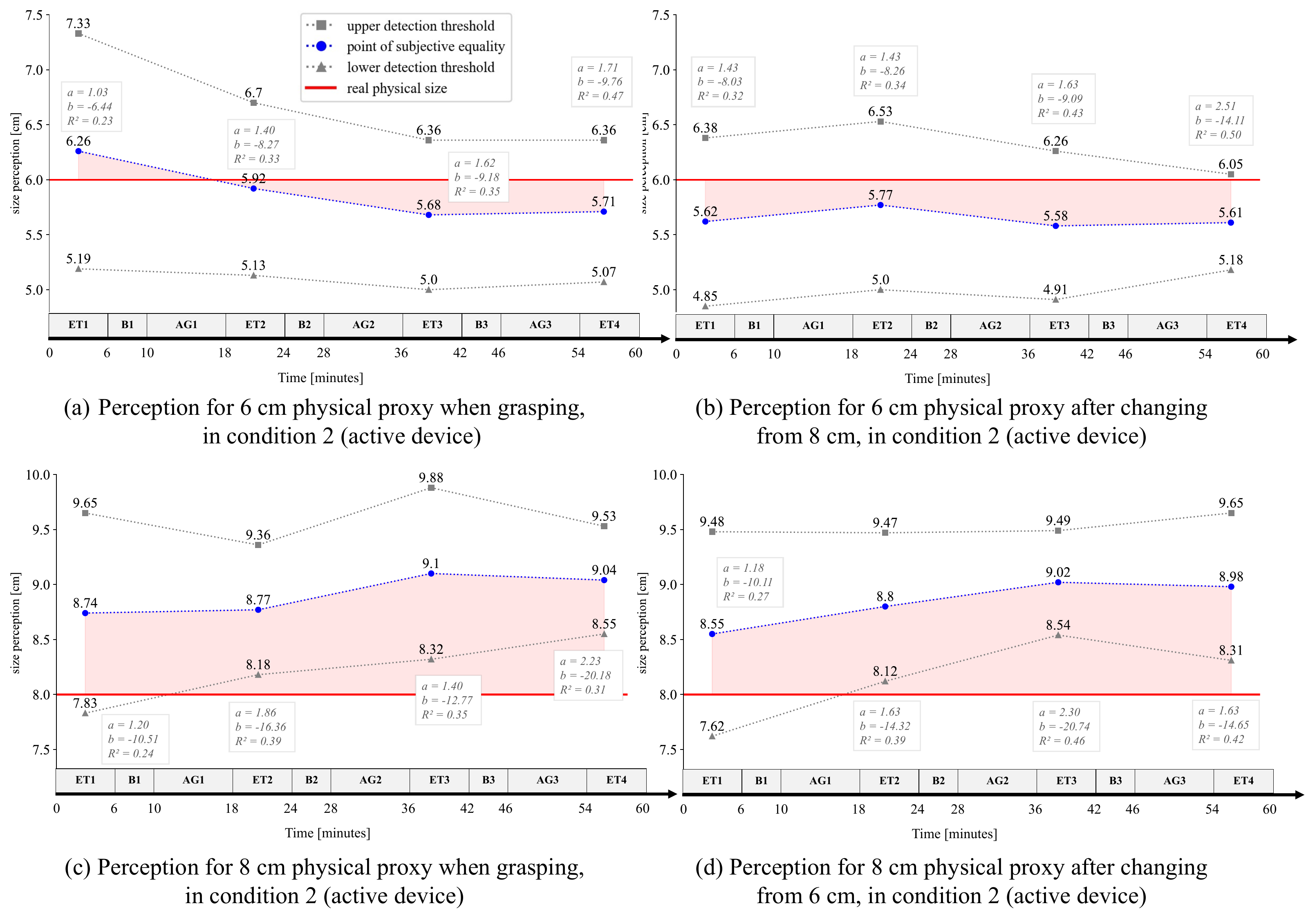}
  \caption{Data analysis of condition 2. Plots are the perception thresholds and PSE of all estimation tasks over time in condition 2. The device performed a size-changing in hand in condition 2 between 6 cm and 8 cm (orders were randomised). And therefore, we have results for both physical sizes before and after size-changing. (a) shows results of 6 cm physical device when grasping (before changing to 8 cm in hand), and (b) shows also 6 cm physical device results but after changing from 8 cm in hand. Similarly, (c) shows results of 8 cm physical device when grasping, before changing to 6 cm, and (d) shows results of 8 cm physical device after changing from 6 cm in hand. Fitting parameters including a, b and McFadden $R^2$ are shown alongside.}
  \Description{The figure presents data analysis for Condition 2, where participants used the size-changing device that switched between 6 cm and 8 cm during the experiment. The plots show perception thresholds and points of subjective equality (PSE) across all estimation tasks over time, separated by physical size and whether participants experienced the size change before or after grasping. Figure (a) shows perception results for the 6 cm physical device when grasping, before it changed to 8 cm. The horizontal represents time across blocks (ET1 to ET4), while the vertical shows size perception in centimetres. The PSE values start at 6.26 cm in ET1 and fluctuate, ending at 5.71 cm in ET4. Dashed lines indicate \camera{upper and lower detection} thresholds. Figure (b) presents perception results for the 6 cm physical device after changing from 8 cm. The PSE values begin at 5.62 cm in ET1, fluctuate through the middle blocks, and end at 5.61 cm in ET4. Figure (c) displays results for the 8 cm physical device when grasping, before it changed to 6 cm. PSE values start at 8.74 cm in ET1 and remain consistently above 8 cm, ending at 9.04 cm in ET4. Figure (d) shows results for the 8 cm physical device after changing from 6 cm. PSE values begin at 7.62 cm in ET1, rise through the middle blocks, and finish at 8.98 cm in ET4.}
  \label{fig:c2}
\end{figure*}

Perception was estimated for two physical sizes (6 cm and 8 cm) at two different moments (immediately after grasping and immediately after the size change), resulting in four scenarios in total presented as four plots within Fig. \ref{fig:c2}. The fittings are all good or strong according to $R^2$ (0.23~0.50). There were 31 (among 20 participants $\times$ 9 virtual sizes $\times$ 2 physical sizes $\times$ 2 answers each lift $\times$ 4 estimation tasks = 2880 questions, 1.07\%) missing answers for condition 2.

Plot (a) is comparable to condition 1, as both present the 6 cm device estimated after grasping. However, unlike condition 1,  the results in condition 2 show a downward trend of size estimation. This indicates that interacting with a size-changing device-even after putting it down and grasping again- affects subsequent size estimation. Initially, the size was overestimated (6.26 cm), though still lower than condition 1 in the first estimation task (6.34 cm). Over time,  the estimated size kept decreasing until it became stable in Estimation Task 4. The size perception changed approximately 5.5 mm (6.26 cm in ET1 - 5.71 cm in ET4).

Plot (b), which shows the 6 cm device after changing from 8 cm, indicates that the 6 cm size was generally underestimated in condition 2. This perception was relatively stable over time.

Plot (c) of 8 cm after grasping before size changes and plot (d) of 8 cm after changing from 6 cm show an increasing perceived size over time and tend to stabilise at the end of the session, similarly to condition 1. The size was largely overestimated for the 8 cm physical size, ranging from 8.55 cm to 9.1 cm.

All four plots show a concentration of perception and a decrease of just noticeable differences (JNDs, half of the difference between the \camera{upper detection thresholds} and \camera{lower detection thresholds}, indicating a range when the users cannot notice the physical and virtual difference).

\subsection{Condition 3 Results}
Results from condition 1 and 2 show that the perception tended to drift over time, further from the real physical size. 
To examine whether a visual priming acts as a form of calibration, reducing or removing this drift, 
we designed condition 3 to offer a potential for visual corrections at different stages of the study.

\begin{figure*}
  \includegraphics[width=0.95\textwidth]{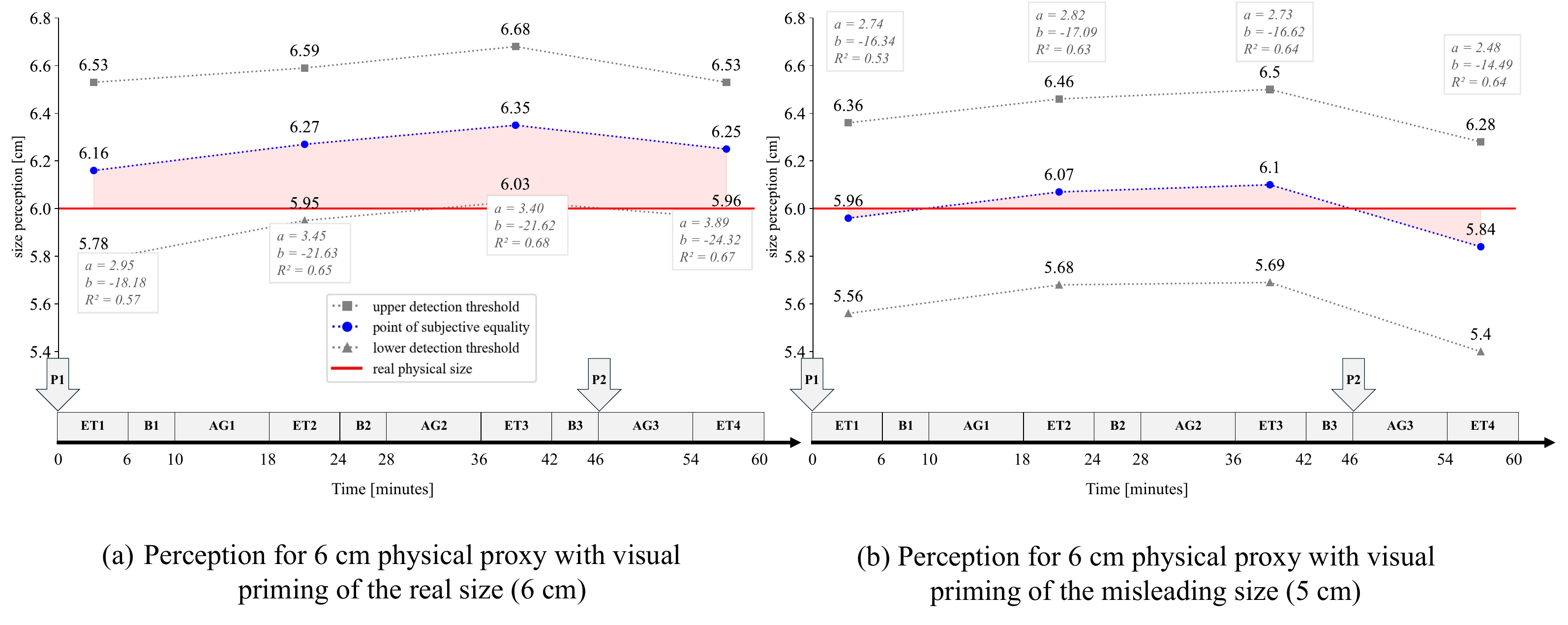}
  \caption{Data analysis of condition 3 and 4. Plots are the perception thresholds and PSE of all estimation tasks over time in condition 3. (a) and (b) are results of condition 3 with the real physical size visual priming (6 cm) and the results of condition 4 with the misleading physical size visual priming (5 cm), respectively. Fitting parameters including a, b and McFadden $R^2$ are shown alongside.}
  \Description{The figure presents data analysis for Conditions 3 and 4, where visual priming was introduced. The plots display perception thresholds and points of subjective equality (PSE) across estimation tasks, with time on the horizontal axis (ET1 through ET4) and perceived size on the vertical axis. Figure (a) shows results for Condition 3, where participants were visually primed with the real size of 6 cm. Perception starts near the real size at 6.16 cm in ET1, increases to 6.35 cm by AG2, and ends at 6.25 cm in ET4. The \camera{upper and lower detection thresholds}, shown as dotted lines, define the perceptual boundaries. Figure (b) presents results for Condition 4, where participants were visually primed with a misleading size of 5 cm. PSE values start at 5.96 cm in ET1, rise slightly above 6 cm during AG2 (6.1 cm), and then drop back to 5.84 cm by ET4.}
  \label{fig:c34}
\end{figure*}

The results of condition 3 are shown in Fig. \ref{fig:c34} (a). Only 4 answers out of 20 participants $\times$ 9 virtual sizes $\times$ 2 times $\times$ 4 estimation tasks = 1440 questions (0.02\%) were missing. The PSE ranges from 6.16 cm to 6.35 cm over time, which is notably smaller than in condition 1 (6.34 to 6.55 cm), where the only difference is the visual prime.

Similar to condition 1, an upward drift of size perception can be observed from estimation task 1 to 3. However, after receiving the second visual priming, prior to the 
final game, the estimation at task 4 dropped back closer to the real physical size (ET4: 6.25 cm).

\subsection{Condition 4 Results}
A misleading visual prime (a 5 cm physical device) was shown to the participants at the beginning of the study and again before the third acclimation game. The real 6 cm device was still used throughout the condition.
The estimated perceptual thresholds and PSEs of condition 4 are shown in Fig. \ref{fig:c34} (b). There were 8 answers out of 20 participants $\times$ 9 virtual sizes $\times$ 2 times $\times$ 4 estimation tasks = 1440 questions (0.06\%) missing. Starting from 5.96 cm in estimation task 1, the perceived size is much closer to the real physical size (6 cm) when the smaller incorrect visual prime was given. Although the perceived size again drifted to be larger over time, the overall perception remained close to the real size. After the misleading small priming was given before acclimation game 3, the size perception dropped dramatically from 6.10 cm to 5.84 cm.

\subsection{NASA-TLX Results}
\begin{figure*}
  \includegraphics[width=0.95\textwidth]{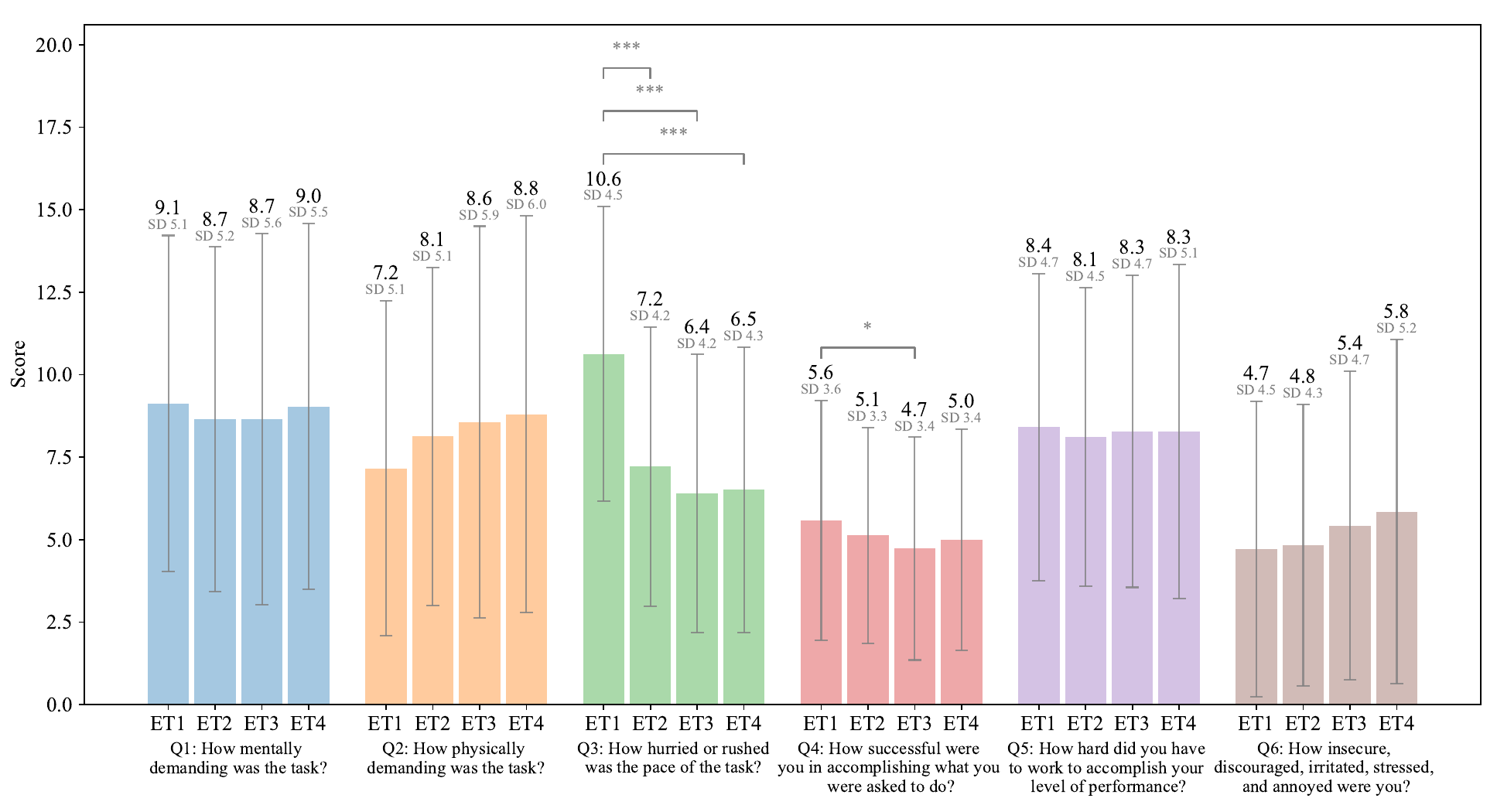}
  \caption{Mean scores and standard deviation (SD) of NASA-TLX questionnaire. ET1-ET4 represent the questionnaires being filled out after estimation task 1 to 4, respectively.}
  \Description{The figure presents the mean scores and standard deviations of the NASA-TLX questionnaire across four estimation tasks (ET1–ET4). The questionnaire measures subjective workload, and each bar cluster corresponds to one of six questions (Q1–Q6). For Q1 (mental demand), scores remain fairly stable across tasks, with means between 8.7 and 9.1 and standard deviations around 5. The Q2 (physical demand) scores show an upward trend, starting at 7.2 in ET1 and rising to 8.8 in ET4. For Q3 (hurried or rushed pace), scores decrease significantly over the tasks: ET1 has the highest mean at 10.6, while ET4 drops to 6.5. Statistical significance is marked with triple asterisks (***), showing that participants felt the pace was less hurried in later tasks compared to the first. In Q4 (success in accomplishing tasks), mean scores are relatively low, ranging from 4.7 to 5.6, suggesting participants felt only moderately successful. A single asterisk (*) indicates a significant difference between ET1 (5.6) and ET2 (5.1). For Q5 (effort to maintain performance), scores remain consistent, averaging around 8.1 to 8.4 across all tasks, with no major fluctuations. Finally, Q6 (insecurity, stress, or irritation) shows a slight upward trend, with scores rising from 4.7 in ET1 to 5.8 in ET4.}
  \label{fig:nasatlx}
\end{figure*}




    The NASA-TLX scores of the 80 participants are summarised in Fig. \ref{fig:nasatlx}. Our intention from the NASA-TLX results was to examine whether effort and mental workload were the mechanisms behind changes in perception.
    The high standard deviations indicate various task load answers from different participants. We analysed the results using a Friedman test and a post-hoc Wilcoxon (Holm-corrected) test.
    The results indicate a significant decrease in temporal demand (Q3) and performance (Q4, note that the decrease in this value means the performance is closer to ``perfect'') over time in different estimation tasks. Although not supported to be significant by the post-hoc test, the mean score and Friedman-test indicate an increase of physical demand (Q2).

\new{For the decreasing temporal pressure (Q3), we interpret that over time the participants were getting used to the task with time limits and, subsequently, not feeling as hurried.} They also seemed to increasingly believe their performance is getting better from the third estimation task (Q4).

%% file: section/5_Discussion.tex
\section{Discussion}

The results of the user study revealed a drift in size perception over time, and a complex interplay between visual and haptic perceptual cues. 
With the abundant results collected, we attempt to explain these observations, present a model of the underlying perceptual mechanisms, and examine their implications for the broader research community.

\subsection{Implications}
\begin{figure*}
  \includegraphics[width=0.95\textwidth]{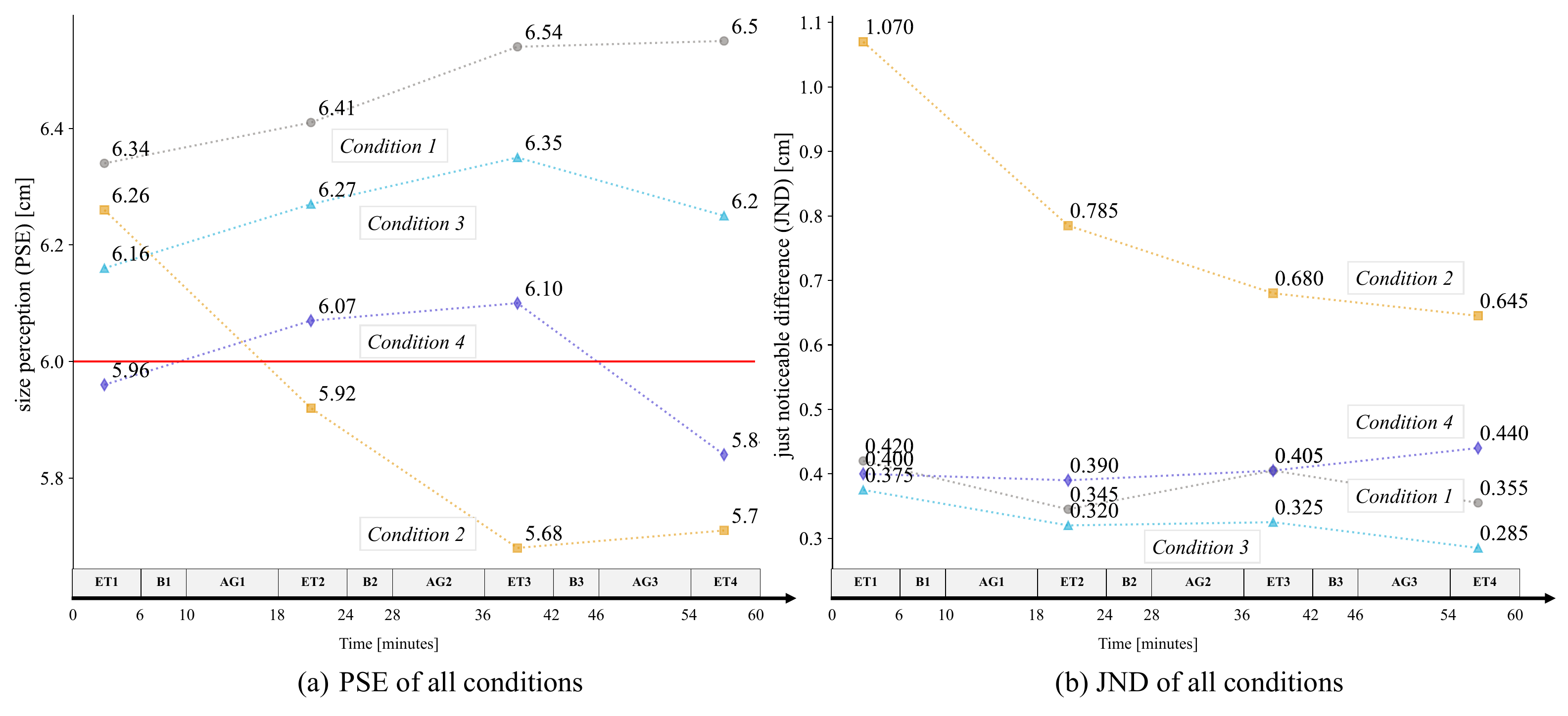}
  \caption{Summary of the results of condition 1-4. Note that the results of condition 2 is represented by the perception of the 6 cm configuration after grasping and before size change. (a) summarises the points of subjective equality and (b) summarises the just noticeable difference by (\camera{upper detection threshold} - \camera{lower detection threshold})/2. The timing of priming of condition 3 and 4 are also marked in the plots.}
  \Description{The figure provides a summary of results across Conditions 1–4, comparing size perception and perceptual sensitivity. Plot (a) shows the Points of Subjective Equality (PSE) for all conditions over time. The horizontal axis represents time (ET1 to ET4), while the vertical axis shows perceived size in centimetres. Condition 1 (no priming, fixed-size device) starts near 6.34 cm and fluctuates upward, peaking at 6.54 cm before stabilizing near 6.5 cm. Condition 2 (size-changing device, 6 cm before size change) starts at 5.92 cm and remains lower than the other conditions, ending at 5.7 cm. Condition 3 (correct priming at 6 cm) remains close to the real size, starting at 6.16 cm and ending at 6.2 cm. Condition 4 (misleading priming at 5 cm) begins at 5.96 cm and stabilizes slightly below the 6 cm line, finishing at 5.8 cm. Plot (b) presents the Just Noticeable Difference (JND) for all conditions, calculated as half the difference between \camera{upper and lower detection thresholds}. Here, the y-axis shows JND in centimetres. Condition 2 shows the highest variability, starting at 1.07 cm and decreasing to 0.645 cm by ET4, indicating participants’ sensitivity improved over time with the active size-changing device. In contrast, Conditions 1, 3, and 4 maintain lower JND values overall, reflecting more stable perceptual sensitivity. Condition 3 (correct priming) shows particularly low JND values (0.325–0.39), suggesting that correct priming improved discrimination ability, while Condition 4 (misleading priming) had slightly higher JNDs (0.405–0.44), reflecting a small but consistent perceptual cost of misleading priming.}
  \label{fig:sum}
\end{figure*}

The PSE and JND results of all conditions are summarised in Fig. \ref{fig:sum}. For the fixed-size objects, all the PSEs drifted to be larger in an almost parallel pattern over time. The estimated size was smaller following visual priming, and became even smaller when following the misleading 5 cm priming. The JNDs of the fixed-size device didn't show any unified continuous pattern. 

The results of condition 2 for the 6 cm physical device, prior to it changing to 8 cm,
differs from the other 6 cm results, revealing how the PSE was increasingly underestimated over time.
The JND was also the largest in this condition, showcasing how swapping sizes led to uncertainty of perception and created a larger illusory area. Over time, this JND became noticeably smaller, indicating that participants gradually became more certain about the object's size. 

\subsubsection{\textbf{Size is overestimated and its perception drifts larger over time for passive objects}}

In the fixed-size conditions, with or without visual priming (condition 1 and 3), the size of the 6 cm physical device was generally overestimated. This overestimation of size has also been identified in previous studies~\cite{bergstrom2019resized, de2019different, jian2025}. For example, in the study of ~\citet{bergstrom2019resized}, perception of 3 physical proxies (3 cm, 6 cm and 9 cm) was estimated 3 times over the course of 
approximately 30 minutes. Therefore, the interaction time for the 6 cm device was about 10 minutes, during which the PSE was estimated to be 6.36 cm, close to the PSE in our first estimation task (6.34 cm).

The \camera{upper and lower detection thresholds} of the previous resized grasping study~\cite{bergstrom2019resized}, however, are 5.4 cm and 7.32 cm for the 6 cm physical object, forming a much larger illusory range than we see in our results in condition 1 (5.92 and 6.76, respectively). This indicates that participants were more confident in their size perception in our study, though 
the basic techniques and procedures employed are similar in both. One potential reason for this is that in these conditions in our study, we only used a 6 cm physical object, where Bergstrom et al.~\cite{bergstrom2019resized} used 3 cm, 6 cm, and 9 cm objects. We hypothesise that our use of a single physical object gave participants more repeated sensory samples, consolidated in working memory, thereby strengthening their internal representation of the object.

Importantly, the overestimation becomes larger over time in fixed-size conditions (condition 1, 3 and 4), and shows a sign of stabilising towards the end of the study, from 39 minutes to 57 minutes (6.54 cm in ET3 to 6.55 cm in ET4 in condition 1). This indicates the size perception drifts towards an asymptote and any perceived size change reduces after a certain time of interacting with the object.

According to the NASA-TLX results, the participants' experience of physical demand increases throughout the study. This could also impact their perception. For example, participants could feel tired and perceive the objects to be heavier and thus larger over time. However, this does not explain the results, as (1) the score change is not significant according to the post-hoc Wilcoxon (Holm-corrected) test, (2) the perception stopped increasing at the end, and (3) the perception drifted towards smaller sizes in condition 2 with the 6 cm device, not following the tired-heavier-larger pattern. 

The perceptual drift can be understood as a sensory phenomenon due to adaptation in cognitive psychology~\cite{jeschke2024humans}. The perception is under real-time adjustment through a closed-loop process with exploration causing the perception to change over time. Beyond perceptual drift as a sensory phenomenon, our results can also be understood in light of cross-modality memorisation. While visual memory and its role in working memory have been extensively studied in HCI to inform design guidelines \cite{morimoto2020nature, hornbaek2025introduction}, cross-modality memorisation has been largely overlooked. Our findings call for further studies in this vein, which could help assess how users build, maintain, and recalibrate their understanding of multimodal interactive systems.


\subsubsection{\textbf{
Grasping an object while it changes size strongly impacts subsequent size perception
}}

In condition 2, we examined size perception both immediately before and after the physical object changes size in each grasp-lift pair. Other than the very first trial, all of the subsequent trials were influenced by the fact that the physical device could change size. We focus on the results of the initial perception after grasping in each grasp-lift pair, rather than on the response after the size change (i.e., we concentrate on the results (a) and (c) in Fig. \ref{fig:c2}).
Although the results immediately after each size change also show smaller JND (for 6 cm and 8 cm configurations) and increasing size over-estimation (for the 8 cm physical device), this is likely a result of complex effects across the virtual size size change, physical size change, visuo-haptic illusion, and a fatigue effect from holding the object in the air without refreshing by putting it down. The perceptual impact and memory of this size change remains to influence subsequent grasp-lift pairs.

Before size change, while the haptic proxy was still 6 cm, the PSE was smaller from the beginning of the study (6.26 cm in ET1), as shown in Fig. \ref{fig:c2} (a) compared with condition 1 (6.34 cm in ET1), and more obviously, drifting towards increasing underestimation over time, which is the opposite pattern to the other conditions with the 6 cm physical device. 

More similarly to conditions 1, 3, and 4, the results of the 8 cm size also drifted to be larger (Fig. \ref{fig:c2}c) and stabilised at the end. The extent of this overestimation was proportionally larger than the overestimation of 6 cm in condition 1. 
This is different from the conclusions of prior work~\cite{bergstrom2019resized}, where sizes were underestimated for larger, 9 cm cubes.

Therefore, we observed that size perception for active, size-changing objects is different to passive, fixed-size objects. 
The frequent changing between physical object sizes influences the visuo-haptic perception. The smaller physical size is perceived as increasingly small, and the larger size is perceived as increasingly large. As this pattern of perception is not observed in prior work with different-sized passive objects, we conclude that the dynamic change of size, and the perceptual cues accumulated during that change, have a lingering impact on subsequent perception of the object when static. In turn, prior motion impacts subsequent perception, but only during contact with an object (at other times, this effect is not seen, so the impact of non-contact motion of perception is, seemingly, attenuated). 

For HCI, this suggests that \textbf{the design of active haptic devices must account not only for the feedback they provide at a given moment, but also for the perceptual aftereffects of dynamic transformations}, since these can bias how users interpret and interact with objects long after the motion has stopped.


\subsubsection{\textbf{Visual priming and its timing influences size perception}}

In condition 3, participants were shown the physical device prior to the first estimation task. This resulted in their estimations being closer to the object's true size, than in condition 1 where they were given no visual prime of the actual device (6.16 cm when primed, in comparison with 6.34 cm when not primed). Though the participants did not see the physical object again for estimation tasks 2 and 3, the impact of that initial visual stimulus remained and their estimations remained lower (though continued to drift slowly upwards, towards 6.35 cm, in comparison to 6.54 cm in condition 1).


Towards the end of the condition, and prior to the final acclimation game and estimation task, the visual prime was given to the participant again. After this, 
the estimated size returned to a more accurate value ($\sim$6.25 cm). This was still not as accurate as the first estimation (6.16 cm). This might result from the 8-minute acclimation game between the visual priming and estimation task 4. If we were to assume that the moment the participants saw the physical device their perception was calibrated to that ground truth (6 cm), then the perceptual drift over time that we see across estimation task 1, 2, and 3, would also explain the drift that occurs across the acclimation game into estimation task 4, yielding the 6.25 cm results. This potentially shows a mathematical relationship between perceptual drift and time.


When a misleading visual prime (5 cm, in condition 4) was presented, the resultant size perception was dragged towards that prime (i.e., initial size perception was 5.96 cm after a 5 cm visual prime, as opposed to 6.16 cm after a 6 cm visual prime). Proportionally, however, this misleading visual prime had a lesser impact on size perception -- the size perception dropped, but did not approach 5 cm. This supports existing theories about optimisation and reliability across different sensory cues ~\cite{ernst2002humans} and how humans do not simply default to trusting vision over haptic cues. 

When the misleading visual prime was presented again, prior to the acclimation game before estimation task 4, the size perception dropped to even 
smaller than the first estimation (5.96 cm), showing that the misleading prime had a larger effect on perception after interaction, than before interaction. The correction and learning between sensory inputs described by Berkeley~\cite{berkeley1709essay} changes at different timings. We posit that this is due to the decreasing reliability of the haptic cues over time, during the interactions, resulting in a subsequent visual cue having a larger impact on sensory integration. 

\subsection{Haptic Drift Model}
\new{It is still unclear how to mathematically describe the drift pattern and why the drift exists in perception. To better describe and interpret the results, we propose a theory to fit and explain the phenomenon. The hypothesis proposed offers one possible perspective and we expect relevant studies to further develop the theory.}

\new{Given the fact that the results drift towards one direction and tend to stabilise, an exponential function is the most direct and simple approach to describe the data, and allows parameters to alter the function to represent the impact of priming.}

\new{We then propose a first-order control system model whose nature is an adaptation process to describe our results. Firstly the exponential mathematical expression can typically describe a first-order control system, and secondly several previous studies in relevant fields, such as haptic retargeting~\cite{10.1145/3491102.3501907, lebrun:tel-04088544} and motor adaptation~\cite{ernst2002humans,yu2024metrics}, proposed similar closed-loop processes/control theories to our theory. The model is intended to be descriptive, as specific parameters may change under different procedures, though we believe the overall flow would remain the same.}

The results show the following patterns that can be observed across more than one condition. 
\begin{itemize}
    \item The visuo-haptic perception drifts towards a certain direction over time (in all conditions).
    \item The haptic drift slows down towards an asymptote and becomes stable (condition 1 and 2)
    \item The initial priming has an impact on subsequent perception throughout the whole study (condition 3 and 4)
    \item The priming in the middle changes the perception directly (condition 3 and 4).
\end{itemize}

First of all, the results of the first condition with 6 cm passive object and no priming can be described as
\begin{equation}
    P = c_1 - c_2\cdot e^{-c_3\cdot t}
\end{equation}
where \textit{P} is the perceived size and \textit{t} is the time since the beginning of the study. The parameters $c_1, c_2$ and $c_3$ describe the initial gap, asymptote, and rate of time accordingly. The exponential approach is chosen because it describes how the perceptual results increase, the rate of increase slows and eventually flattens out, 
approaching a constant value.

Note that the fit can be transformed into a first-order control system expression mathematically. Therefore, we propose a \new{visuo-haptic perception} drift model, with the expression:

\begin{equation}
    P(t) = P_\infty + (P_0-P_\infty)e^{-\frac{t}{\tau}}
\end{equation}

where $P_0$ is the initial value of the model, showing the initial perception at the beginning of interaction, $P_\infty$ is the steady state value, meaning the final perception formed after infinite time of interaction and $\tau$ is the time constant ($\tau$ is the time when the value is about 63\% of the way to $P_\infty$ and $3\tau$ is the time when the value is 95\% of the way to $P_\infty$).

\begin{figure*}
  \includegraphics[width=0.95\textwidth]{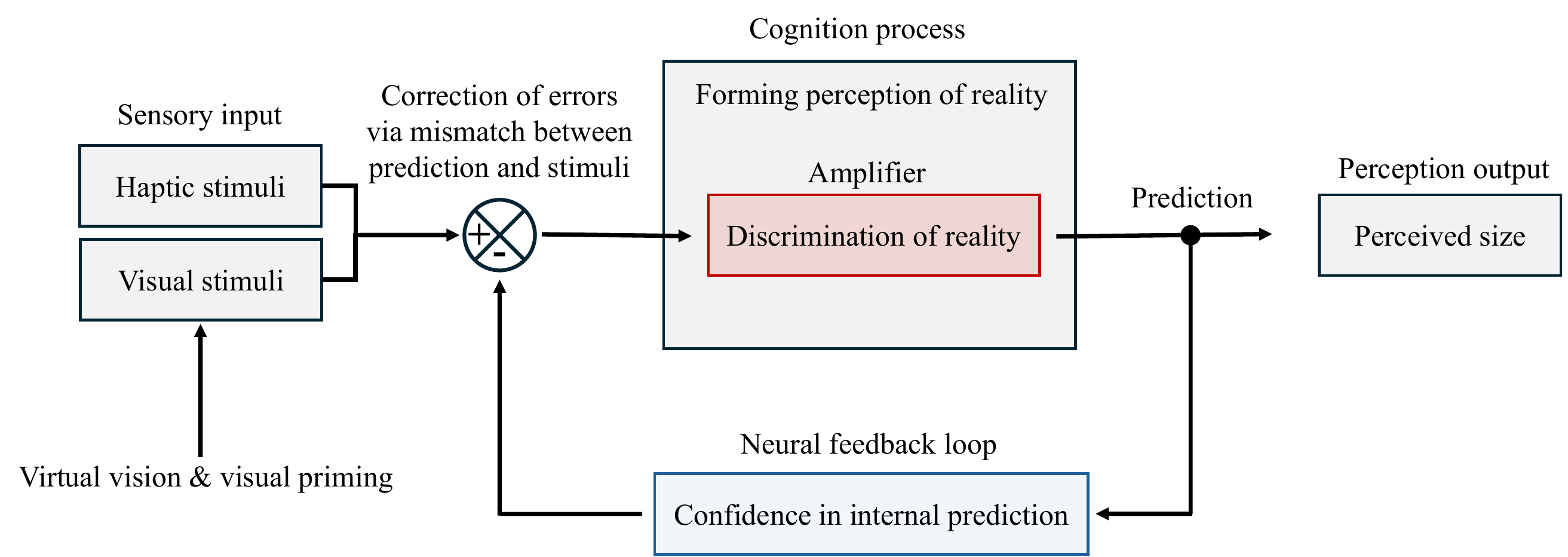}
  \caption{A model to describe the perception process as a first-order control system. The sensory input corrects the perceptual errors through a feedback loop over time, and discrimination in the cognitive process results in an amplified perception at last.}
  \Description{The figure illustrates a conceptual model of perception framed as a first-order control system, showing how sensory inputs, cognitive processes, and feedback mechanisms interact to shape perceived size. On the left, sensory input consists of haptic stimuli and visual stimuli, which include both virtual vision and visual priming. These inputs are processed together, and discrepancies between perception and actual stimuli are corrected through a mismatch error correction mechanism. The corrected signals are then passed into the cognition process, where the perception of reality is formed. Within this process, an amplifier highlights the role of discrimination of reality, which emphasises how the brain magnifies perceptual judgments when resolving sensory conflicts. From cognition, a prediction is generated, leading to the perception output, represented here as the perceived size of the object. At the same time, the prediction feeds into a neural feedback loop, where confidence in internal prediction is reinforced. This feedback loop interacts with the error correction mechanism, ensuring ongoing adjustments between expected and actual sensory information.}
  \label{fig:model}
\end{figure*}

When modelling perception, there are usually three components involved: \textit{sensory information}, \textit{expectation}, and \textit{attention}~\cite{hornbaek2025introduction}. 
A first-order control system is characterised by a differential equation relating inputs and outputs. 
If we assume human perception shares the same pattern with the control system, the haptic and visual stimuli are the sensory input (\textit{sensory information}) and the perceived size is the output of the system (estimated by our forced choice task) as shown in Fig. \ref{fig:model}. Humans dynamically perceive the haptic stimuli across the one-hour duration of the study, and adjust the output based on their best estimation, derived from their perception
(\textit{prediction/expectation}). Over time, this estimation stabilises. 

During the active process of the \new{visuo-haptic perception}, people adopt exploratory behaviours to gain sensory stimulus, to inform their understanding of objects and the surrounding environment. When these different signals are processed, however, noise is introduced into our understanding (which we term 'discrimination of reality' in our model). This has been observed previously, such as with resized grasping~\cite{bergstrom2019resized, de2019different, jian2025}, retargeting~\cite{azmandian2016haptic}, and the rubber hand illusion~\cite{botvinick1998rubber}, where the perception didn't match the reality accurately. As a result of this noise, humans amplify the momentary sensory cues in the output perception. Their outcome understanding, as a prediction of reality, is repeatedly assessed against new incoming sensory cues in an attempt to reduce the errors. 


These predictions formed in the perceptual system are partially corrected by incoming sensory signals via recurrent feedback loops, through which humans maximise the information and attempt to reduce perceptual errors. However, the extent to which predictive signals are weighted within the brain remains unclear. Confidence in internal predictions appears to bias this weighting of reliability, thereby constraining the accuracy of error correction. As a result, residual error persists even after the system reaches stability. This confidence in comparison is also related with \textit{attention} and sometimes is noisy and lossy. Bayesian models for sensory integration are used to explain the weights of prior information and prediction~\cite{kording2004bayesian, berniker2011bayesian}. Over time, without continuous calibration, perceptual experience becomes amplified, leading to a measurable mismatch between sensory input and perceptual output.



\begin{figure}
  \includegraphics[width=0.475\textwidth]{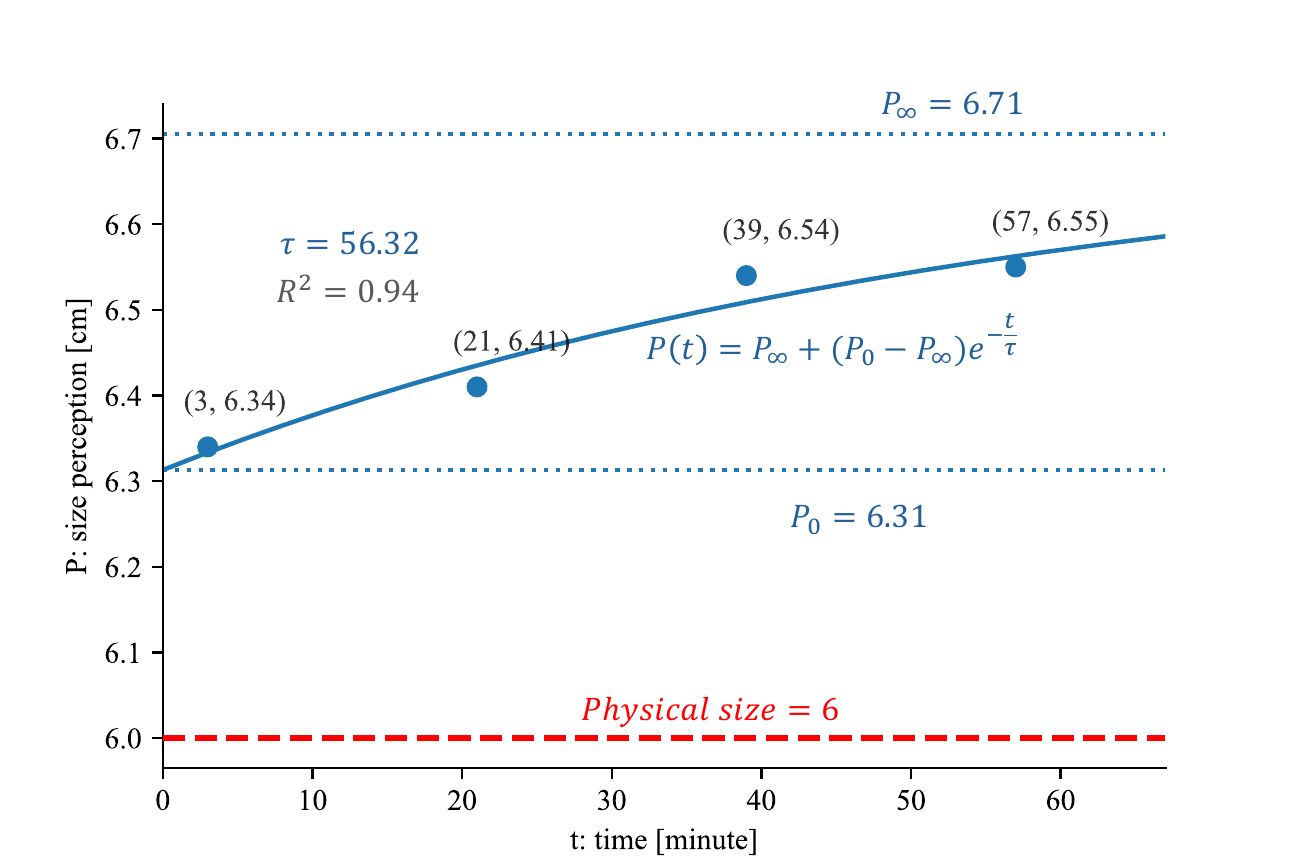}
  \caption{Haptic drift model of fixed-size device without any visual priming (condition 1).}
  \Description{The figure shows the haptic drift model for the fixed-size proxy in condition 1, where no visual priming was used. The vertical axis represents size perception in centimetres, and the horizontal axis represents time in minutes across a 60 minute period. The points mark the observed perceptions at different times. At 3 minutes the perceived size was 6.34 centimetres, at 21 minutes it was 6.41 centimetres, at 39 minutes it was 6.54 centimetres, and at 57 minutes it was 6.55 centimetres. These values are all above the actual physical size, which is shown as a dashed red line at 6 centimetres. A curve is fitted to the data, showing how perception increased gradually over time. The initial perceived size was about 6.31 centimetres, and over the course of the study it drifted upward toward about 6.71 centimetres. The time constant of this drift was about 56 minutes, and the model fit was very strong with an R-squared value of 0.94.}
  \label{fig:m1}
\end{figure}

The result of fitting the data points of the 6 cm physical device without visual priming is shown in Fig. \ref{fig:m1} and the fitting parameters are $c_1 = 6.71$, $c_2 = 0.39$ and $c_3 = 0.02$, accordingly, with $R^2 = 0.94$ indicating a good fit. This model reveals an
initial perception $P_0=6.31$, steady state perception (the asymptote) $P_\infty=6.71$, and time constant $\tau = 56.32$ (Fig.\ref{fig:m1}). This means the initial perception at the beginning of interaction, although not estimated in the experiment, is 6.31 cm according to the model. After enough time, the perception will finally approach 6.71 cm. The time constant $\tau$ shows that after 56.32 minutes of interaction, the perception is around 63\% of the way to $P_\infty$. When the interaction time is longer than $3\tau$ the perception will be 95\% of $P_\infty$.

Therefore, while it is almost impossible to estimate the perception over such a long duration with user studies, the model predicts the starting point and ending state of the perception. More importantly, this reveals that human cognition acts in a similar way to a control system - a procedure of initial guess, correction, and approaching a steady state.

\begin{figure*}
  \includegraphics[width=0.95\textwidth]{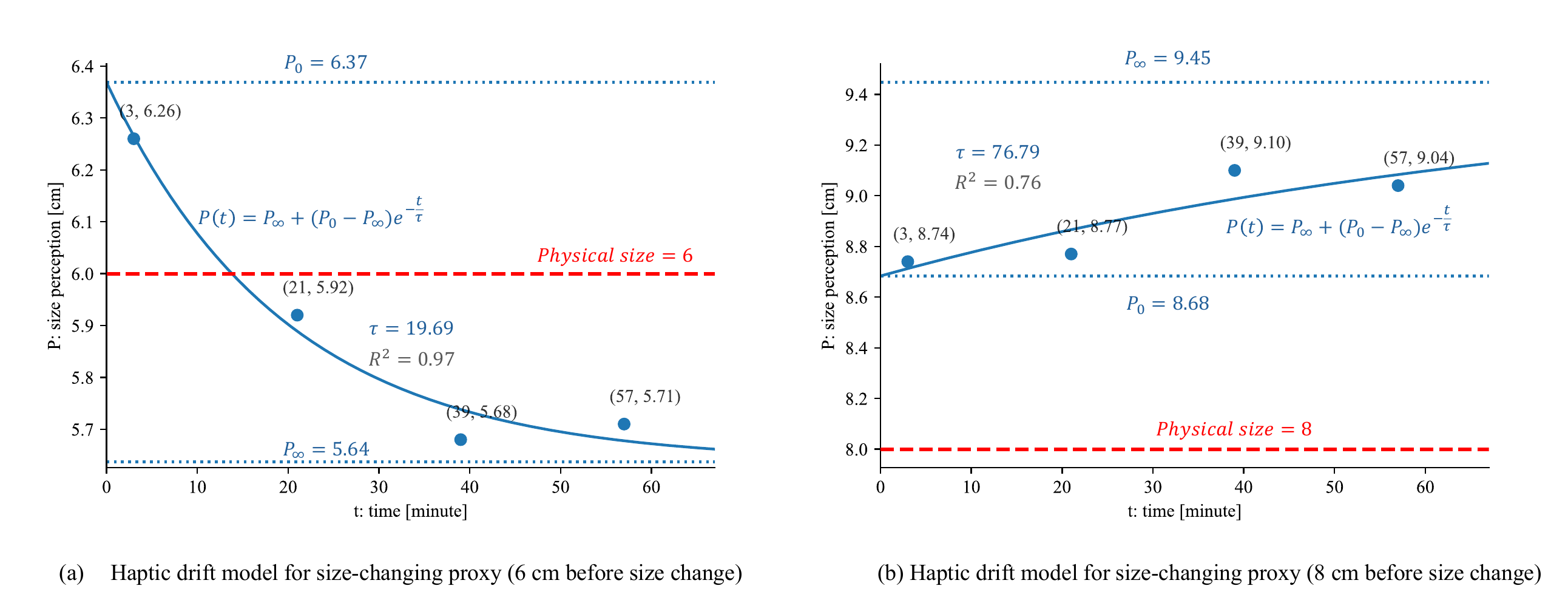}
  \caption{Haptic drift model of size-changing device (condition 2). (a) shows the perception of 6 cm configuration before size change and (b) shows the perception of 8 cm configuration before size change.}
  \Description{The figure presents the haptic drift model for the size-changing proxy in condition 2, showing perception for both the 6 centimetre and 8 centimetre configurations before size change. In plot (a), the vertical axis shows perceived size in centimetres and the horizontal axis shows time in minutes. Data points mark the perceived size of the 6 centimetre configuration before the proxy changed size. At 3 minutes the perception was 6.26 centimetres, at 21 minutes it was 5.92 centimetres, at 39 minutes it was 5.68 centimetres, and at 57 minutes it was 5.71 centimetres. A fitted curve shows that perception started higher, around 6.37 centimetres, and gradually decreased toward 5.64 centimetres. The drift had a time constant of about 20 minutes, and the model fit was very strong with an R squared of 0.97. In panel (b), the same type of analysis is shown for the 8 centimetre configuration before size change. The data points show perceived sizes of 8.74 centimetres at 3 minutes, 8.77 centimetres at 21 minutes, 9.10 centimetres at 39 minutes, and 9.04 centimetres at 57 minutes. The fitted curve shows perception starting lower, at about 8.68 centimetres, and drifting upward toward 9.45 centimetres. The drift had a longer time constant of about 77 minutes, and the model fit was moderate with an R squared of 0.76.}
  \label{fig:m2}
\end{figure*}

Similarly, the size-changing device perception can be fitted to the model as shown in Fig. \ref{fig:m2} with $R^2$ being 0.97 and 0.76 for the small and larger configurations accordingly. Again, we focus on the perception after grasping and before size change, because the perception after size changing is a complex procedure with multiple changing sensory stimuli.

For the 6 cm configuration, the initial perception $P_0$ is 6.37 cm, larger than the physical size. The steady state perception $P_\infty$, however, drops to 5.64 cm. The time constant $\tau$ is 19.69 minutes, showing that the perception drift is 95\% done within 1 hour (3$\tau$).

The initial perception $P_0$ (8.68 cm) and steady state perception $P_\infty$ (9.45 cm) are both much larger than the physical size 8 cm in the larger configuration of the device. However, this procedure takes longer time to complete with $\tau = 76.79$ minutes and $3\tau = 230.37$ minutes.

\begin{figure*}
  \includegraphics[width=0.95\textwidth]{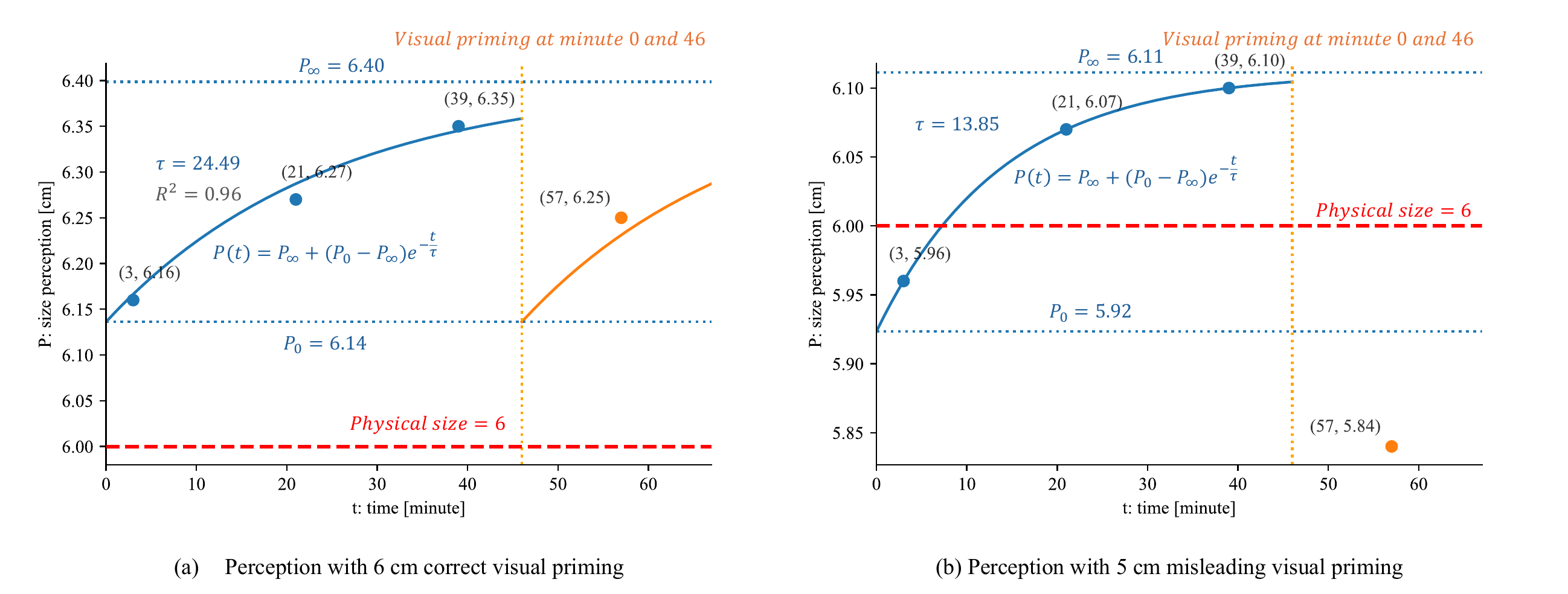}
  \caption{Haptic drift model of fixed-size device with visual priming (condition 3 and 4). (a) shows the perception with the real physical size (6 cm) visual priming and (b) shows the perception with the misleading physical size (5 cm) visual priming.}
  \Description{The figure shows the haptic drift model for the fixed-size proxy under visual priming conditions. In plot (a), the vertical axis shows size perception in centimetres and the horizontal axis shows time in minutes. Data points indicate perceived size of the 6 centimetre proxy over the study. At 3 minutes perception was 6.16 centimetres, at 21 minutes it was 6.27 centimetres, at 39 minutes it was 6.35 centimetres, and at 57 minutes it was 6.25 centimetres. A fitted curve shows perception starting near 6.14 centimetres and drifting toward 6.40 centimetres, with a time constant of about 24 minutes and a strong model fit of R squared 0.96. The visual priming events happen at the start (0 minutes) and again at 46 minutes. In plot (b), the same analysis is applied for misleading visual priming at 5 centimetres. Perceptions began at 5.96 centimetres at 3 minutes, rose slightly to 6.07 centimetres at 21 minutes, reached 6.10 centimetres at 39 minutes, and then dropped to 5.84 centimetres at 57 minutes. The fitted curve indicates perception starting at about 5.92 centimetres and drifting toward 6.11 centimetres, with a time constant of about 14 minutes and a moderate fit of R squared 0.82. Visual priming events at 0 and 46 minutes are again indicated.}
  \label{fig:m34}
\end{figure*}

In condition 3, we presented the correct visual prime again at minute 46 and, thus, we assume the perception is corrected to $P_0$ again from this point (through a process akin to sensory dead-reckoning). From this assumption, the perception is modelled as shown in Fig. \ref{fig:m34} (a) with $R^2 = 0.96$. The initial perception is $P_0 = 6.14$ cm, but then drifts towards $P_\infty = 6.40$ cm. At minute 46, \new{we propose another theoretical hypothesis that the perception is refreshed to $P_0$ and the procedure of drift starts over again.} With an initial correct visual priming, both the initial perception (6.14 cm) and steady state perception (6.40 cm) are more accurate compared with non-priming condition 1 (initial perception 6.31 cm, steady state perception 6.71 cm). Also the time constant $\tau$ (24.49 minutes) is much smaller in value than condition 1 ($\tau = 56.32$), indicating the perception gets to steady state much faster with visual priming.

As the second visual priming was presented in condition 4 with the misleading stimuli (5 cm), a sudden drop in perception was observed and the mechanism is complex. Thus, the final data point of condition 4 is excluded from the model. With the other 3 data points, the initial perception, steady state perception and time constant are calculated to be $P_0=5.92$ cm, $P_\infty = 6.11$ cm and $\tau=13.85$ minutes. With the misleading smaller visual priming of 5 cm, the perception will still eventually drift to 6.11 cm to be larger than the physical size 6 cm, and the procedure is fast with $\tau=13.85$ minutes and $3\tau =41.55$ minutes.

By applying our \new{visuo-haptic perception} drift model, whose nature is mathematically a first-order control system, we interpret the procedure of human \new{visuo-haptic perception} and how it changes in different priming conditions, for different types of devices. The model can also calculate and predict perception unable, or not available any more, to be estimated.

Although many previous studies on perception and motor control indicate an improvement in accuracy over time\cite{lee2018moving, hornbaek2025introduction}, the model shows the accuracy in our task only gets stabilised but is lower in general.

While it's hard to quantify the impact of visual priming in the model, the model indicates a ''calibration'' function of visual priming during interaction, which may relate to previous theories of visual and haptic integration models, that a maximum-likelihood integrator decides which sensory information should take domain in perception~\cite{ernst2002humans}.

The model aligns with the perception theory in HCI with \textit{sensory information}, \textit{expectation} and \textit{attention}~\cite{hornbaek2025introduction}.
Our model also aligns with historical and contemporary accounts of cross-sensory calibration. 
Berkeley’s Essay Towards a New Theory of Vision (1709) \cite{berkeley1709essay} argued that vision is calibrated by touch; a view that has since been revisited in multisensory research \cite{gori2012visual, briscoe2015action}, and shows that visual input often anchors haptic estimates. 
In our experiments, intermittent visual priming acted precisely as such a calibration signal, resetting accumulated haptic drift. 
Within our study, a correct visual prime operates as an external corrective input that reduces error and stabilises the feedback loop. 
Moreover, because recalibrated estimates persisted after priming, we speculate that working memory retains this ``reset'' state, delaying the re-emergence of drift. This highlights how cross-modal calibration and working memory jointly contribute to the dynamics of perceptual stability in tangible interaction.

\subsection{Application} 

\textbf{Perception changes over time.} As shown in all the results, perception changes during interaction. Our proposed model reveals how this perception changes and, thus, what users perceive about objects they interact with over time. This model gives interaction designers clearer insights, without having to repeatedly run psychometric tests to understand perception change during use. 


\textbf{Passive objects represent larger virtual objects.} As shown in Fig. \ref{fig:c1} (b), a passive object of 6 cm is generally overestimated in interaction. After about 21 minutes of estimation, the downscaling perceptual threshold (6.06 cm) will be larger than the physical object itself (6 cm). Given prior claims around how experience degrades when objects fall outside of this range ~\cite{jian2025}, this means that, for example, the physical object could no longer be used to represent itself in a virtual reality experience. This effect can be mitigated with visual priming. 

\textbf{Active devices impact perception differently to multiple passive devices.} Many previous studies and designs assumed that passive and active devices shared the same patterns (e.g. ~\citet{jian2025} applied the illusion spaces model to both passive and active devices in discussion). However, our results warn the research and design community to be more cautious on their perceptual conclusions, as the active devices yield meaningfully different results to the fixed-size objects.
The size change provides an obvious contrasting stimulus between configurations. After a while of interaction (approximately minute 39 to 57 including breaks) the small configurations (6 cm) represents even smaller virtual objects (5.68~5.71 cm) and accordingly, the large configuration (8 cm) represents even larger virtual objects (9.04~9.1 cm). This shows that in simple active devices, smaller dynamic size changes result in larger perceived size changes.
In more complex designs with multiple configurations, the size should be carefully designed to reach the target effect.

\textbf{Seeing the physical object affects its ability to represent objects.} Many works in the research and design community ignored the influence of revealing the object to the participants during studies. While in most applications, the users would know the visual appearance of the object, the results estimated when the objects were not shown to the participants can be inaccurate in these scenarios. However, the results can be corrected, or at least predicted closer to reality, by utilising the haptic drift model and applying visual priming parameters. 

Presenting the visual priming seems to calibrate the results and reset the sensory errors. This means that well-designed breaks, for example, during which participants can see the physical device, have a meaningful impact on their ongoing perception. 
This is even more important when the appearance of the device is designed to mislead the users into believing something else, because showing the misleading stimuli again after some interaction has an even larger effect.

The results also suggest that potentially concealing or hiding the mechanism of the physical device may affect user's perception, leaving them with an incomplete or misleading visual reference for the object they are interacting with. 

\subsection{Future Works}

In our study a \new{visuo-haptic perception} drift was observed in VR. It is still to investigated how this drift happens, and if this drift applies to a much larger range of interaction, even including the most common touch and grasp interactions of all objects in the physical world. \new{We propose a first-order control system theory and an exponential mathematical model. Future related studies in human-computer interaction, psychology, and cognitive science are needed to further develop the theory.}

While a prolonged interaction causes drifts in perception, it is worth exploring if there can be approaches to accelerate this procedure for effective interaction (e.g., training and education), or to slow down the drift to reduce perception errors. 

\new{As this work primarily concentrated on the fundamental factors with a specific setup, several future works are required to generalise the conclusion and theory. For example, \textbf{interaction effect} (visual priming and active device) or other factors may affect this process. \textbf{Individual perception} might also be different from our population conclusion we obtained in this study. However, it is still a challenge to enhance the generalisability in these perceptual findings to become effective both for the population and the individuals. The \textbf{grasping behaviour and type} is also a common limitation in the visuo-haptic research community. In our study we examined the moment right after grasping and lifting, and a certain grasp type. The impacts of the timing and types are still underexplored. While it's obviously tedious work to estimate on all timings and grasp types in user studies, some biomechanical/neurological interpretations behind them may form some theories, patterns or models for the community to expand our conclusions in the future. Similarly, all the \textbf{interactive objects properties} can not be included in a single study. While we employed a specific device, it's potentially predictable how the results may change when the properties differ. For example, \citet{bergstrom2019resized} and \citet{jian2025} modelled the perceptions when the properties change. Yet it is not verified if all conclusions, including those in this paper, will follow the proposed model in previous studies and thus need future verifications.}

\camera{While the visualisation of the users' hands impacts tangible interactions~\cite{jorg2020virtual,10049656}, technical challenges persist for precisely tracking the whole-hand movement. Therefore, only simplified indicators were rendered in our user study to help grasp and avoid any tracking marker occlusion or inaccurate hand rendering which would have largely impacted the results. Although the absolute values of the perceived size may slightly change with different hand rendering techniques in practice, we believe the major conclusions, including the drift patterns and influence of priming still hold true, which may be verified in future work.}



Future work could also investigate how entrenched mental representations of everyday objects—formed as sensory traces consolidate into long-term memory, shape current perception in VR. 
This raises important HCI questions about how novel haptic devices can either exploit or must work against such perceptual baselines when simulating familiar objects. 


While our study focused on size perception, it is also worth investigating whether similar forms of haptic drift occur in other object properties, such as weight, shape, texture, temperature, and stiffness. 
Exploring these dimensions would help determine whether drift is a generalisable feature of \new{visuo-haptic perception} or one specific to size. 
If perceptual drift reflects a more fundamental neurological process of forming reality in the brain, then comparable shifts might emerge across nearly all sensory modalities, not only in haptics. 
For HCI, this suggests that prolonged interaction with haptic devices could gradually alter how users perceive familiar material properties, potentially enhancing immersion but also introducing subtle mismatches when devices fail to recalibrate perception effectively. 
Studying these broader visuo-haptic properties would therefore deepen our understanding of perceptual stability and guide the design of haptic systems that remain convincing over extended use.

%% file: section/6_Conclusion.tex
\section{Conclusion}

Understanding how people form and adjust visuo-haptic perception is critical for HCI, as it shapes the design of tangible interfaces and haptic systems. 
To investigate this process, we conducted a user study across four conditions, using forced choice tasks to examine how perception evolves over time and how it is influenced by visual priming.
Our findings show that size perception drifts in a consistent direction before flattening out during prolonged interaction, and that both correct and misleading visual primes significantly alter how participants interpret the object’s property. 
Based on these results, we modelled visuo-haptic perception as a closed feedback loop: sensory signals are integrated, calibrated against prior expectations, and gradually stabilise into cross-model perceptual representations.
Taken together, these findings highlight that visuo-haptic perception is not static but dynamically shaped by time, calibration, and cross-modal interaction, offering both new opportunities and important challenges for the design of tangible interactive technologies.




%% file: base.bib
@inproceedings{10.1145/3472749.3474782,
 address = {New York, NY, USA},
 author = {Gonzalez, Eric J and Ofek, Eyal and Gonzalez-Franco, Mar and Sinclair, Mike},
 booktitle = {The 34th Annual ACM Symposium on User Interface Software and Technology},
 doi = {10.1145/3472749.3474782},
 isbn = {9781450386357},
 location = {Virtual Event, USA},
 numpages = {11},
 pages = {732–742},
 publisher = {Association for Computing Machinery},
 series = {UIST '21},
 title = {X-Rings: A Hand-mounted 360° Shape Display for Grasping in Virtual Reality},
 url = {https://doi.org/10.1145/3472749.3474782},
 year = {2021}
}

@inproceedings{azmandian2016haptic,
 author = {Azmandian, Mahdi and Hancock, Mark and Benko, Hrvoje and Ofek, Eyal and Wilson, Andrew D},
 booktitle = {Proceedings of the 2016 chi conference on human factors in computing systems},
 doi = {10.1145/2858036.2858226},
 pages = {1968--1979},
 title = {Haptic retargeting: Dynamic repurposing of passive haptics for enhanced virtual reality experiences},
 year = {2016}
}

@inproceedings{ban2012modifying,
 author = {Ban, Yuki and Kajinami, Takashi and Narumi, Takuji and Tanikawa, Tomohiro and Hirose, Michitaka},
 booktitle = {2012 IEEE Haptics Symposium (HAPTICS)},
 doi = {10.1109/HAPTIC.2012.6183793},
 organization = {IEEE},
 pages = {211--216},
 title = {Modifying an identified curved surface shape using pseudo-haptic effect},
 year = {2012}
}

@inproceedings{ban2014displaying,
 author = {Ban, Yuki and Narumi, Takuji and Tanikawa, Tomohiro and Hirose, Michitaka},
 booktitle = {Proceedings of the 20th ACM Symposium on Virtual Reality Software and Technology},
 doi = {10.1145/2671015.2671028},
 pages = {191--196},
 title = {Displaying shapes with various types of surfaces using visuo-haptic interaction},
 year = {2014}
}

@inproceedings{bergstrom2019resized,
 author = {Bergstr{\"o}m, Joanna and Mottelson, Aske and Knibbe, Jarrod},
 booktitle = {Proceedings of the 32nd Annual ACM Symposium on User Interface Software and Technology},
 doi = {10.1145/3332165.3347939},
 pages = {1175--1183},
 title = {Resized grasping in vr: Estimating thresholds for object discrimination},
 year = {2019}
}

@incollection{bickmann2019haptic,
 author = {Bickmann, Raoul and Tran, Celine and Ruesch, Ninja and Wolf, Katrin},
 booktitle = {Proceedings of Mensch und Computer 2019},
 doi = {10.1145/3340764.3344459},
 pages = {565--569},
 title = {Haptic illusion glove: a glove for illusionary touch feedback when grasping virtual objects},
 year = {2019}
}

@inproceedings{brahimaj2023cross,
 author = {Brahimaj, Detjon and Berthaut, Florent and Giraud, Fr{\'e}d{\'e}ric and Semail, Betty},
 booktitle = {Proceedings of the 34th Conference on l'Interaction Humain-Machine},
 pages = {1--12},
 title = {Cross-modal interaction of stereoscopy, surface deformation and tactile feedback on the perception of texture roughness in an active touch condition: Interaction intermodale de la st{\'e}r{\'e}oscopie, de la d{\'e}formation de surface et de la retour tactile sur la perception de la rugosit{\'e} de la texture dans un {\'e}tat tactile actif},
 year = {2023},
doi = {https://doi.org/10.1145/3583961.3583967}

}

@inproceedings{choi2016wolverine,
 author = {Choi, Inrak and Hawkes, Elliot W and Christensen, David L and Ploch, Christopher J and Follmer, Sean},
 booktitle = {2016 IEEE/RSJ International Conference on Intelligent Robots and Systems (IROS)},
 doi = {10.1109/iros.2016.7759169},
 organization = {IEEE},
 pages = {986--993},
 title = {Wolverine: A wearable haptic interface for grasping in virtual reality},
 year = {2016}
}

@inproceedings{choi2018claw,
 author = {Choi, Inrak and Ofek, Eyal and Benko, Hrvoje and Sinclair, Mike and Holz, Christian},
 booktitle = {Proceedings of the 2018 CHI conference on human factors in computing systems},
 doi = {10.1145/3173574.3174228},
 pages = {1--13},
 title = {Claw: A multifunctional handheld haptic controller for grasping, touching, and triggering in virtual reality},
 year = {2018}
}

@inproceedings{clarence2021unscripted,
 author = {Clarence, Aldrich and Knibbe, Jarrod and Cordeil, Maxime and Wybrow, Michael},
 booktitle = {2021 IEEE Virtual Reality and 3D User Interfaces (VR)},
 doi = {10.1109/vr50410.2021.00036},
 organization = {IEEE},
 pages = {150--159},
 title = {Unscripted retargeting: Reach prediction for haptic retargeting in virtual reality},
 year = {2021}
}

@inproceedings{clarence2022investigating,
 author = {Clarence, Aldrich and Knibbe, Jarrod and Cordeil, Maxime and Wybrow, Michael},
 booktitle = {2022 IEEE International Symposium on Mixed and Augmented Reality (ISMAR)},
 doi = {10.1109/ismar55827.2022.00078},
 organization = {IEEE},
 pages = {612--621},
 title = {Investigating the effect of direction on the limits of haptic retargeting},
 year = {2022}
}

@inproceedings{clarence2024stacked,
 author = {Clarence, Aldrich and Knibbe, Jarrod and Cordeil, Maxime and Wybrow, Michael},
 booktitle = {Proceedings of the CHI Conference on Human Factors in Computing Systems},
 doi = {10.1145/3613904.3642228},
 pages = {1--13},
 title = {Stacked Retargeting: Combining Redirected Walking and Hand Redirection to Expand Haptic Retargeting's Coverage},
 year = {2024}
}

@inproceedings{de2019different,
 author = {de Tinguy, Xavier and Pacchierotti, Claudio and Emily, Mathieu and Chevalier, Mathilde and Guignardat, Aur{\'e}lie and Guillaudeux, Morgan and Six, Chlo{\'e} and L{\'e}cuyer, Anatole and Marchal, Maud},
 booktitle = {2019 ieee world haptics conference (whc)},
 doi = {10.1109/whc.2019.8816164},
 organization = {IEEE},
 pages = {580--585},
 title = {How different tangible and virtual objects can be while still feeling the same?},
 year = {2019}
}

@article{feix2015grasp,
 author = {Feix, Thomas and Romero, Javier and Schmiedmayer, Heinz-Bodo and Dollar, Aaron M and Kragic, Danica},
 doi = {10.1109/thms.2015.2470657},
 journal = {IEEE Transactions on human-machine systems},
 number = {1},
 pages = {66--77},
 publisher = {IEEE},
 title = {The grasp taxonomy of human grasp types},
 volume = {46},
 year = {2015}
}

@inproceedings{follmer2013inform,
 author = {Follmer, Sean and Leithinger, Daniel and Olwal, Alex and Hogge, Akimitsu and Ishii, Hiroshi},
 booktitle = {Uist},
 doi = {10.1145/2501988.2502032},
 number = {10},
 organization = {Citeseer},
 pages = {2501--988},
 title = {inFORM: dynamic physical affordances and constraints through shape and object actuation.},
 volume = {13},
 year = {2013}
}

@inproceedings{gonzalez2019investigating,
 author = {Gonzalez, Eric J and Follmer, Sean},
 booktitle = {Proceedings of the 25th ACM Symposium on Virtual Reality Software and Technology},
 doi = {10.1145/3359996.3364248},
 pages = {1--5},
 title = {Investigating the detection of bimanual haptic retargeting in virtual reality},
 year = {2019}
}

@inproceedings{gonzalez2020reach+,
 author = {Gonzalez, Eric J and Abtahi, Parastoo and Follmer, Sean},
 booktitle = {Proceedings of the 33rd Annual ACM Symposium on User Interface Software and Technology},
 doi = {10.1145/3379337.3415870},
 pages = {236--248},
 title = {Reach+ extending the reachability of encountered-type haptics devices through dynamic redirection in vr},
 year = {2020}
}

@inproceedings{kim2024big,
 author = {Kim, Myung Jin and Ofek, Eyal and Pahud, Michel and Sinclair, Mike J and Bianchi, Andrea},
 booktitle = {Proceedings of the CHI Conference on Human Factors in Computing Systems},
 doi = {10.1145/3613904.3642254},
 pages = {1--15},
 title = {Big or Small, It’s All in Your Head: Visuo-Haptic Illusion of Size-Change Using Finger-Repositioning},
 year = {2024}
}

@inproceedings{kohli2010redirected,
 author = {Kohli, Luv},
 booktitle = {2010 IEEE Symposium on 3D User Interfaces (3DUI)},
 doi = {10.1109/3dui.2010.5444703},
 organization = {IEEE},
 pages = {129--130},
 title = {Redirected touching: Warping space to remap passive haptics},
 year = {2010}
}

@phdthesis{lebrun:tel-04088544,
 author = {Lebrun, Flavien},
 doi = {10.1145/3649792.3649803},
 hal_id = {tel-04088544},
 hal_version = {v1},
 month = {December},
 number = {2022SORUS525},
 school = {{Sorbonne Universit{\'e}}},
 title = {{Study of Visuo-Haptic Illusions in Virtual Reality : understanding and Predicting Illusion Detection}},
 type = {Theses},
 url = {https://theses.hal.science/tel-04088544},
 year = {2022}
}

@inproceedings{lecuyer2000pseudo,
 author = {L{\'e}cuyer, Anatole and Coquillart, Sabine and Kheddar, Abderrahmane and Richard, Paul and Coiffet, Philippe},
 booktitle = {Proceedings IEEE Virtual Reality 2000 (Cat. No. 00CB37048)},
 doi = {10.1109/vr.2000.840369},
 organization = {IEEE},
 pages = {83--90},
 title = {Pseudo-haptic feedback: Can isometric input devices simulate force feedback?},
 year = {2000}
}

@article{sanz2013elastic,
 author = {Sanz, Ferran Argelaguet and J{\'a}uregui, David Antonio G{\'o}mez and Marchal, Maud and L{\'e}cuyer, Anatole},
 doi = {10.1145/2501599},
 journal = {ACM Transactions on Applied Perception},
 number = {3},
 pages = {17--1},
 title = {Elastic images: Perceiving local elasticity of images through a novel pseudo-haptic deformation effect},
 volume = {10},
 year = {2013}
}

@inproceedings{siu2018shapeshift,
 author = {Siu, Alexa F and Gonzalez, Eric J and Yuan, Shenli and Ginsberg, Jason B and Follmer, Sean},
 booktitle = {Proceedings of the 2018 CHI Conference on Human Factors in Computing Systems},
 doi = {10.1145/3173574.3173865},
 pages = {1--13},
 title = {Shapeshift: 2D spatial manipulation and self-actuation of tabletop shape displays for tangible and haptic interaction},
 year = {2018}
}

@inproceedings{turchet2010influence,
 author = {Turchet, Luca and Marchal, Maud and L{\'e}cuyer, Anatole and Nordahl, Rolf and Serafin, Stefania},
 booktitle = {Proceedings of the 17th ACM Symposium on Virtual Reality Software and Technology},
 doi = {10.1145/1889863.1889893},
 pages = {139--142},
 title = {Influence of auditory and visual feedback for perceiving walking over bumps and holes in desktop VR},
 year = {2010}
}

@inproceedings{ulan2024,
 author = {Kelesbekov, Ulan and Marini, Gabriele  and Zhongyi, Bai and Johal, Wafa and Velloso, Eduardo and Knibbe, Jarrod},
 booktitle = {2024 IEEE International Symposium on Mixed and Augmented Reality (ISMAR)},
 doi = {10.1109/ISMAR62088.2024.00045},
 organization = {IEEE},
 title = {Stuet: Dual Stewart Platforms for Pinch Grasping Objects in VR},
 year = {2024}
}

@inproceedings{yang2018vr,
 author = {Yang, Jackie and Horii, Hiroshi and Thayer, Alexander and Ballagas, Rafael},
 booktitle = {Proceedings of the 31st Annual ACM Symposium on User Interface Software and Technology},
 doi = {10.1145/3242587.3242643},
 pages = {889--899},
 title = {VR Grabbers: Ungrounded haptic retargeting for precision grabbing tools},
 year = {2018}
}

@inproceedings{zenner2019estimating,
 author = {Zenner, Andr{\'e} and Kr{\"u}ger, Antonio},
 booktitle = {2019 IEEE Conference on Virtual Reality and 3D User Interfaces (VR)},
 doi = {10.1109/vr.2019.8798143},
 organization = {IEEE},
 pages = {47--55},
 title = {Estimating detection thresholds for desktop-scale hand redirection in virtual reality},
 year = {2019}
}

@inproceedings{zhao2018functional,
 author = {Zhao, Yiwei and Follmer, Sean},
 booktitle = {Proceedings of the 2018 CHI Conference on Human Factors in Computing Systems},
 doi = {10.1145/3173574.3174118},
 pages = {1--12},
 title = {A functional optimization based approach for continuous 3d retargeted touch of arbitrary, complex boundaries in haptic virtual reality},
 year = {2018}
}

@inproceedings{jian2025,
  bibtex_show = {true},
  author = {Zhang, Jian and Knibbe, Jarrod and Johal, Wafa},
  title = {Illusion Spaces in VR: The Interplay Between Size and Taper Angle Perception in Grasping},
  booktitle = {Proceedings of the CHI Conference on Human Factors in Computing Systems (CHI '25)},
  year = {2025},
  location = {Yokohama, Japan},
  doi = {10.1145/3706598.3714162},
  acmisbn = {979-8-4007-1394-1/25/04},
  publisher = {ACM},
  selected = {yes}
}

@article{liu2021quantitative,
  title={Quantitative investigation of hand grasp functionality: Thumb grasping behavior adapting to different object shapes, sizes, and relative positions},
  author={Liu, Yuan and Zeng, Bo and Jiang, Li and Liu, Hong and Ming, Dong},
  journal={Applied Bionics and Biomechanics},
  volume={2021},
  number={1},
  pages={2640422},
  year={2021},
doi= {10.1155/2021/2640422},
  publisher={Wiley Online Library}
}

@article{napier1956prehensile,
  title={The prehensile movements of the human hand},
  author={Napier, John R},
  journal={The Journal of Bone \& Joint Surgery British Volume},
  volume={38},
  number={4},
  pages={902--913},
  year={1956},
doi={10.1302/0301-620X.38B4.902},
  publisher={Bone \& Joint}
}

@inproceedings{10.1145/3290605.3300923,
author = {Zhu, Kening and Chen, Taizhou and Han, Feng and Wu, Yi-Shiun},
title = {HapTwist: Creating Interactive Haptic Proxies in Virtual Reality Using Low-cost Twistable Artefacts},
year = {2019},
isbn = {9781450359702},
publisher = {Association for Computing Machinery},
address = {New York, NY, USA},
url = {https://doi.org/10.1145/3290605.3300923},
doi = {10.1145/3290605.3300923},
booktitle = {Proceedings of the 2019 CHI Conference on Human Factors in Computing Systems},
pages = {1–13},
numpages = {13},
location = {Glasgow, Scotland Uk},
series = {CHI '19}
}

@article{botvinick1998rubber,
  title={Rubber hands ‘feel’touch that eyes see},
  author={Botvinick, Matthew and Cohen, Jonathan},
  journal={Nature},
  volume={391},
  number={6669},
  pages={756--756},
  year={1998},
  doi = {https://doi.org/10.1038/35784},
  publisher={Nature Publishing Group UK London}
}

@article{krakauer2019motor,
  title={Motor learning},
  author={Krakauer, John W and Hadjiosif, Alkis M and Xu, Jing and Wong, Aaron L and Haith, Adrian M},
  journal={Comprehensive physiology},
  volume={9},
  number={2},
  pages={613--663},
  year={2019},
  doi = {10.1002/j.2040-4603.2019.tb00069.x},
  publisher={Wiley Online Library}
}

@article{ernst2002humans,
  title={Humans integrate visual and haptic information in a statistically optimal fashion},
  author={Ernst, Marc O and Banks, Martin S},
  journal={Nature},
  volume={415},
  pages={429--433},
  year={2002},
  publisher={Nature Publishing Group},
  doi={10.1038/415429a}
}

@article{ERNST2004162,
title = {Merging the senses into a robust percept},
journal = {Trends in Cognitive Sciences},
volume = {8},
number = {4},
pages = {162-169},
year = {2004},
issn = {1364-6613},
doi = {https://doi.org/10.1016/j.tics.2004.02.002},
url = {https://www.sciencedirect.com/science/article/pii/S1364661304000385},
author = {Marc O. Ernst and Heinrich H. Bülthoff}
}

@article{fetsch2013bridging,
  title={Bridging the gap between theories of sensory cue integration and the physiology of multisensory neurons},
  author={Fetsch, Christopher R and DeAngelis, Gregory C and Angelaki, Dora E},
  journal={Nature Reviews Neuroscience},
  volume={14},
  number={6},
  pages={429--442},
  year={2013},
  publisher={Nature Publishing Group},
  doi={10.1038/nrn3503}
}

@book{berkeley1709essay,
  title={An essay towards a new theory of vision},
  author={Berkeley, George},
  year={1709},
  publisher={IndyPublish. com}
}

@article{gori2012visual,
  title={Visual size perception and haptic calibration during development},
  author={Gori, Monica and Giuliana, Luana and Sandini, Giulio and Burr, David},
  journal={Developmental science},
  volume={15},
  number={6},
  pages={854--862},
  year={2012},
  doi = {https://doi.org/10.1111/j.1467-7687.2012.01183.x},
  publisher={Wiley Online Library}
}

@article{morimoto2020nature,
  title={The nature of haptic working memory capacity and its relation to visual working memory},
  author={Morimoto, Taku},
  journal={Multisensory Research},
  volume={33},
  number={8},
  pages={837--864},
  year={2020},
doi = {https://doi.org/10.1163/22134808-bja10007},
  publisher={Brill}
}

@article{jeschke2024humans,
  title={Humans flexibly use visual priors to optimize their haptic exploratory behavior},
  author={Jeschke, Michaela and Zoeller, Aaron C and Drewing, Knut},
  journal={Scientific Reports},
  volume={14},
  number={1},
  pages={14906},
  year={2024},
doi = {https://doi.org/10.1038/s41598-024-65958-6},
  publisher={Nature Publishing Group UK London}
}

@book{clark2024experience,
  title={The experience machine: How our minds predict and shape reality},
  author={Clark, Andy},
  year={2024},
  publisher={Random House}
}

@article{briscoe2015action,
  title={Action-based theories of perception},
  author={Briscoe, Robert and Grush, Rick and Springle, Alison},
  year={2015}
}

@book{hornbaek2025introduction,
  title={Introduction to Human-Computer Interaction},
  author={Hornb{\ae}k, Kasper and Kristensson, Per Ola and Oulasvirta, Antti},
  year={2025},
  publisher={Oxford University Press}
}

@inproceedings{lee2018moving,
  title={Moving target selection: A cue integration model},
  author={Lee, Byungjoo and Kim, Sunjun and Oulasvirta, Antti and Lee, Jong-In and Park, Eunji},
  booktitle={Proceedings of the 2018 CHI Conference on Human Factors in Computing Systems},
  pages={1--12},
doi = {https://doi.org/10.1145/3173574.3173804},
  year={2018}
}

@inproceedings{yu2024metrics,
  title={Metrics of motor learning for analyzing movement mapping in virtual reality},
  author={Yu, Difeng and Cibulskis, Mantas and Mortensen, Erik Skjoldan and Christensen, Mark Schram and Bergstr{\"o}m, Joanna},
  booktitle={Proceedings of the 2024 CHI Conference on Human Factors in Computing Systems},
  pages={1--18},
doi = {https://doi.org/10.1145/3613904.3642354},
  year={2024}
}

@article{johansson1992sensory,
  title={Sensory-motor coordination during grasping and manipulative actions},
  author={Johansson, Roland S and Cole, Kelly J},
  journal={Current opinion in neurobiology},
  volume={2},
  number={6},
  pages={815--823},
  year={1992},
doi = {https://doi.org/10.1016/0959-4388(92)90139-C},
  publisher={Elsevier}
}

@incollection{hart1988development,
  title={Development of NASA-TLX (Task Load Index): Results of empirical and theoretical research},
  author={Hart, Sandra G and Staveland, Lowell E},
  booktitle={Advances in psychology},
  volume={52},
  pages={139--183},
  year={1988},
  DOI = {https://doi.org/10.1016/S0166-4115(08)62386-9},
  publisher={Elsevier}
}

@article{berniker2011bayesian,
  title={Bayesian approaches to sensory integration for motor control},
  author={Berniker, Max and Kording, Konrad},
  journal={Wiley Interdisciplinary Reviews: Cognitive Science},
  volume={2},
  number={4},
  pages={419--428},
  year={2011},
doi = {https://doi.org/10.1002/wcs.125},
  publisher={Wiley Online Library}
}

@article{kording2004bayesian,
  title={Bayesian integration in sensorimotor learning},
  author={K{\"o}rding, Konrad P and Wolpert, Daniel M},
  journal={Nature},
  volume={427},
  number={6971},
  pages={244--247},
  year={2004},
doi ={https://doi.org/10.1038/nature02169},
  publisher={Nature Publishing Group UK London}
}

@ARTICLE{10049656,
  author={Venkatakrishnan, Roshan and Venkatakrishnan, Rohith and Raveendranath, Balagopal and Pagano, Christopher C. and Robb, Andrew C. and Lin, Wen-Chieh and Babu, Sabarish V.},
  journal={IEEE Transactions on Visualization and Computer Graphics}, 
  title={How Virtual Hand Representations Affect the Perceptions of Dynamic Affordances in Virtual Reality}, 
  year={2023},
  volume={29},
  number={5},
  pages={2258-2268},
  doi={10.1109/TVCG.2023.3247041}}

@incollection{jorg2020virtual,
  title={Virtual hands in VR: Motion capture, synthesis, and perception},
  author={J{\"o}rg, Sophie and Ye, Yuting and Mueller, Franziska and Neff, Michael and Zordan, Victor},
  booktitle={SIGGRAPH Asia 2020 Courses},
  pages={1--32},
doi={https://doi.org/10.1145/3415263.3419155},
  year={2020}
}

@ARTICLE{9580906,
  author={Zhang, Li and He, Weiping and Hu, Yupeng and Wang, Shuxia and Bai, Huidong and Billinghurst, Mark},
  journal={IEEE Access}, 
  title={Using the Visuo-Haptic Illusion to Perceive and Manipulate Different Virtual Objects in Augmented Reality}, 
  year={2021},
  volume={9},
  number={},
  pages={143782-143794},
  doi={10.1109/ACCESS.2021.3121390}}

@inproceedings{10.1145/3491102.3501907,
author = {Gonzalez, Eric J and Chase, Elyse D. Z. and Kotipalli, Pramod and Follmer, Sean},
title = {A Model Predictive Control Approach for Reach Redirection in Virtual Reality},
year = {2022},
isbn = {9781450391573},
publisher = {Association for Computing Machinery},
address = {New York, NY, USA},
url = {https://doi.org/10.1145/3491102.3501907},
doi = {10.1145/3491102.3501907},
booktitle = {Proceedings of the 2022 CHI Conference on Human Factors in Computing Systems},
articleno = {39},
numpages = {15},
location = {New Orleans, LA, USA},
series = {CHI '22}
}

@INPROCEEDINGS{Zhang2025ThumbShift,
  author={Zhang, Jian and Buckingham, Gavin and Johal, Wafa and Knibbe, Jarrod},
  booktitle={2025 IEEE International Symposium on Mixed and Augmented Reality (ISMAR)}, 
  title={ThumbShift: Modulating Perceived Object Properties Through Dynamic Thumb Repositioning}, 
  year={2025},
  volume={},
  number={},
  pages={571-581},
  doi={10.1109/ISMAR67309.2025.00067}}
